%% file: _MAIN.tex
\documentclass[11pt,a4paper]{article}

\usepackage[T1]{fontenc}
\usepackage[utf8]{inputenc}
\usepackage{lmodern}
\usepackage[margin=2.5cm]{geometry}
\usepackage{amsmath,amssymb,amsthm}
\usepackage{graphicx}
\usepackage{booktabs}
\usepackage{adjustbox}
\usepackage{pdflscape}
\usepackage{float}
\usepackage{algorithm}
\usepackage{algpseudocode}
\usepackage[font=small,labelfont=bf,skip=6pt]{caption}
\usepackage[super]{cite}
\usepackage{authblk}
\usepackage{xcolor}
\usepackage[hypertexnames=false]{hyperref}
\hypersetup{colorlinks=true,linkcolor=blue!60!black,citecolor=blue!60!black,urlcolor=blue!60!black}

\theoremstyle{definition}
\newtheorem{definition}{Definition}
\theoremstyle{plain}
\newtheorem{proposition}{Proposition}

\newcommand{\A}{\tilde{A}}
\newcommand{\W}{\mathcal{W}}
\newcommand{\Cl}{\mathcal{C}}
\newcommand{\tbl}[2]{\caption{#1}#2}
\newcommand{\alttext}[1]{}

\title{Community Detection in Fuzzy Social Networks through Fuzzy Co-Membership Relations over $\alpha$-Cut Realizations}

\author[1,2]{Annamaria Porreca}
\author[3]{Fabrizio Maturo\thanks{Corresponding author: \href{mailto:fabrizio.maturo@unimercatorum.it}{fabrizio.maturo@unimercatorum.it}}}
\affil[1]{Department for the Promotion of Human Science and Quality of Life, San Raffaele University, Rome, Italy}
\affil[2]{Unit of Clinical and Molecular Epidemiology, IRCCS San Raffaele Roma, Rome, Italy}
\affil[3]{Faculty of Technological and Innovation Sciences, Department of Economics, Statistics and Business, Universitas Mercatorum, Piazza Mattei 10, 00186 Rome, Italy}
\date{}

\begin{document}

\maketitle

\begin{abstract}
In fuzzy social networks the ties between actors are triangular fuzzy numbers, and community detection applied to their cores ignores how the partition depends on their vagueness. We propose a fuzzy co-membership relation obtained by running the Louvain algorithm on crisp networks compatible with the alpha-cuts of the ties and integrating the proportions of co-assignment over alpha. The relation is positive semidefinite; from it we derive a consensus partition, the number of groups, membership degrees and an assignment ambiguity index. In simulations from stochastic block models the relation does not yield better partitions than symmetric perturbations of the cores of the same size. Its ambiguity index is more closely associated with the structural position of the actors and identifies actors split between two groups when the vagueness is concentrated on them; when the asymmetry of the fuzzy ties points towards the latent tie strength, its co-membership degrees have a lower Brier score. The interval-weighted Louvain algorithm estimates the number of communities more accurately. In two departmental collaboration networks the observed ambiguity is lower than under permuted ratings, while the partitions depend on the symmetrization of directed ratings and on the coding of non-response.
\end{abstract}


\section{Introduction}\label{sec:intro}

In fuzzy social networks the ties between actors are fuzzy numbers rather than crisp weights. Porreca et al.\cite{porreca2025} introduced directed fuzzy social networks with ties represented by triangular fuzzy numbers (TFNs), collected through a mouse-tracking questionnaire inspired by Calcagn\`i and Lombardi,\cite{calcagni2014} and extended centrality indices to this setting. This paper addresses the mesoscopic level: the detection of groups of actors that are more closely tied among themselves than with the rest of the network.\cite{fortunato2010}

Fuzzy approaches to community detection mostly introduce fuzziness in the output, through overlapping or graded memberships, while ties are crisp;\cite{su2014,nino2019,golsefid2015,naderipour2020,luo2020,cangemi2025} Nepusz et al.\cite{nepusz2008} combine fuzzy memberships with a bridgeness index, and other contributions add soft information on the nodes.\cite{gutierrez2022,elmoussaoui2025} The closest contribution is the interval-weighted Louvain algorithm of Alves et al.,\cite{alves2024} which extends modularity to interval-valued ties and returns one crisp partition. A TFN carries more than one interval: its $\alpha$-cuts form a nested family, and a natural question is how the community structure changes as the vagueness of the ties is taken into account at different levels.

We explore, for each level $\alpha$, the crisp networks compatible with the $\alpha$-cuts of the ties, partition each of them with a modularity-based algorithm, and integrate the proportions of co-assignment over $\alpha$. The result is a fuzzy co-membership relation between actors, from which we derive a consensus partition, the number of groups, membership degrees and an assignment ambiguity index. The construction belongs to consensus and evidence-accumulation clustering.\cite{monti2003,fred2005,lancichinetti2012} Perturbing edge weights and recording how often nodes are clustered together is an established way to identify unstable nodes and to assess the robustness of communities,\cite{gfeller2005,karrer2008} but there the size of the perturbation is chosen by the analyst. Here the perturbations are generated by the fuzzy ties: their size, their distribution across ties and their asymmetry are those stated by the respondents, and the nested $\alpha$-cuts provide a family of perturbation sizes.

Whether this adds anything to a perturbation of the cores of the same size is the main question of the simulation study. We compare the proposed relation with the consensus of repeated runs on the cores and with two procedures that perturb the cores symmetrically by the same amount as the $\alpha$-cuts, with a common or a tie-specific scale. Two further designs ask when the vagueness is informative: one concentrates it on actors that split their ties between two blocks, the other generates asymmetric TFNs whose longer side points, or does not point, towards the latent tie strength. The aim is not a better partition, which the simulations do not support, but a relation whose degrees describe how the community structure depends on the stated vagueness of the ties.

Section~\ref{sec:method} presents the method, Section~\ref{sec:sim} the simulation study, Section~\ref{sec:real} an application to two departmental collaboration networks, and Section~\ref{sec:disc} concludes. The proof and detailed simulation tables are in Appendices A and B.

\section{Methodology}\label{sec:method}

\subsection{Fuzzy social networks and $\alpha$-cuts}\label{sec:framework}

Let $V=\{v_1,\dots,v_n\}$ be a set of actors. A directed fuzzy social network\cite{porreca2025} is a pair $\tilde G=(V,\A)$ with $\A_{ij}=(l_{ij},m_{ij},r_{ij})$, $0\le l_{ij}\le m_{ij}\le r_{ij}\le1$, and $(0,0,0)$ for absent ties; $L$, $M$ and $R$ collect the three parameters. The $\alpha$-cut of a tie is $[W^L_{\alpha,ij},W^U_{\alpha,ij}]$ with
\begin{equation}\label{eq:cut}
W^L_\alpha=L+\alpha(M-L),\qquad W^U_\alpha=R-\alpha(R-M),\qquad\alpha\in[0,1],
\end{equation}
and the crisp networks compatible with the ties at level $\alpha$ form the box $\W_\alpha=\{W:W^L_\alpha\le W\le W^U_\alpha\}$. The boxes are nested, $\W_{\alpha'}\subseteq\W_\alpha$ for $\alpha\le\alpha'$, and $\W_1=\{M\}$.

Community detection is performed on an undirected projection $W^s_{ij}=\Phi(W_{ij},W_{ji})$, with $\Phi$ non-decreasing in both arguments. The relation of interest, collaboration, is symmetric, and $\A_{ij}$ and $\A_{ji}$ are two reports of the same relation; in the application only respondents report ties, so the ties sent by non-respondents are all null. We use $\Phi=\max$, which declares a tie when at least one of the two actors reports it and keeps in the analysis the actors who are only rated by others. Since $\max$ is continuous and non-decreasing, the extension principle gives the $\alpha$-cuts of the projected tie as $[\max(W^L_{\alpha,ij},W^L_{\alpha,ji}),\max(W^U_{\alpha,ij},W^U_{\alpha,ji})]$, which contain the projection of every network in $\W_\alpha$. A modularity analysis of the directed ratings would treat the actors who send no ties as sinks, an artifact of non-response. The effect of $\Phi$ is examined in Section~\ref{sec:real}.

Best-path connectivity, the max-min transitive closure of the ties used by Porreca et al.,\cite{porreca2025} has an ultrametric complement and induces the partitions of single linkage:\cite{zadeh1971,gower1969} one strong tie between two groups merges them. With a strict t-norm such as the product\cite{klement2000} the ultrametric is avoided, but path-based connectivity still measures reachability, not membership of a group denser in ties than expected. We therefore build on modularity\cite{newman2004} and on the Louvain algorithm.\cite{blondel2008}

\subsection{Fuzzy co-membership}\label{sec:proposed}

Let $\Cl$ be the Louvain algorithm, which maps a weighted undirected network to a partition of $V$, heuristically seeks to maximize modularity and determines the number of communities. It is randomized through the order in which nodes are visited.

\begin{definition}\label{def:comemb}
For $\alpha\in[0,1]$ let $W^{(1)}_\alpha,\dots,W^{(B)}_\alpha\in\W_\alpha$, $B>3$, be realizations with $W^{(1)}_\alpha=W^L_\alpha$, $W^{(2)}_\alpha=W^U_\alpha$, $W^{(3)}_\alpha=(W^L_\alpha+W^U_\alpha)/2$ and $W^{(b)}_{\alpha,ij}=W^L_{\alpha,ij}+U^{(b)}_{ij}(W^U_{\alpha,ij}-W^L_{\alpha,ij})$, $U^{(b)}_{ij}\sim\mathcal U(0,1)$ independent, for $b\ge4$. Let $P^{(b)}_\alpha=\Cl\big(\Phi(W^{(b)}_\alpha,W^{(b)\top}_\alpha)\big)$, each from an independent run, and let $w_b=\lambda/3$ for $b\le3$ and $w_b=(1-\lambda)/(B-3)$ for $b\ge4$, $\lambda\in(0,1)$. The level-wise co-membership and the fuzzy co-membership relation are
\begin{equation}\label{eq:mu}
\pi_{ij}(\alpha)=\sum_{b=1}^B w_b\,\mathbb 1\{i\sim_{P^{(b)}_\alpha}j\},\qquad
\mu_{ij}=\int_0^1\pi_{ij}(\alpha)\,d\alpha,\qquad\mu_{ii}=1,
\end{equation}
where $i\sim_Pj$ means that $i$ and $j$ share a community of $P$; the integral is computed by the trapezoidal rule on a grid $0=\alpha_1<\dots<\alpha_J=1$ with weights $\omega_1,\dots,\omega_J$.
\end{definition}

The realizations explore the $\alpha$-cut without assuming a probability model for the ties, and $\mu$ depends on the exploration law. The three deterministic realizations, with total weight $\lambda$, include the extreme configurations of the box; the uniform draws, with weight $1-\lambda$, perturb the ties independently. Since the weights do not depend on $B$, increasing $B$ reduces the Monte Carlo error without changing the law. We use $\lambda=3/25$ and $B=25$, so that all realizations have weight $1/25$. At $\alpha=1$ all realizations equal the cores and $\pi_{ij}(1)$ reflects only the randomness of the algorithm. The relation $\mu$ is a frequency of co-assignment over the explored networks, not the probability that $i$ and $j$ belong to the same latent community.

\begin{definition}\label{def:control}
The consensus of the cores, $\mu^{\mathrm{core}}$ (Consensus-core), is obtained from Eq.~(\ref{eq:mu}) with every realization equal to $M$ and the same $JB$ runs and weights. Let $\bar\Delta_{ij}$ be the mean absolute deviation $|W_{ij}-m_{ij}|$ of the FCo-L realizations of an observed tie, computed exactly from its $\alpha$-cuts and the weights $w_b$ and $\omega_g$. For $U\sim\mathcal U(0,1)$ the clipped perturbation $\min\{1,\max\{0,m+\delta(2U-1)\}\}$ has mean absolute deviation $\rho(0,m)=0$ and, for $\delta>0$, $\rho(\delta,m)=\{f(\delta,1-m)+f(\delta,m)\}/(2\delta)$, with $f(\delta,t)=\delta^2/2$ for $\delta\le t$ and $f(\delta,t)=t\delta-t^2/2$ otherwise; $\rho$ is continuous and increasing in $\delta$. Perturb-core uses $JB$ equally weighted realizations $W^{(b)}_{ij}=\min\{1,\max\{0,m_{ij}+\delta(2U^{(b)}_{ij}-1)\}\}$ for observed ties and $0$ otherwise, with a common $\delta$ solving $\mathrm{mean}_{ij}\,\rho(\delta,m_{ij})=\mathrm{mean}_{ij}\,\bar\Delta_{ij}$; Perturb-tie uses a tie-specific $\delta_{ij}$ solving $\rho(\delta_{ij},m_{ij})=\bar\Delta_{ij}$.
\end{definition}

The control relation captures the instability of the algorithm on the cores. Perturb-core matches the average size of the FCo-L perturbation, Perturb-tie matches it tie by tie, so that FCo-L differs from Perturb-tie only in the shape of the exploration: the asymmetry of the TFNs, the mixture of perturbation sizes over $\alpha$ and the extreme realizations. With independent runs the variance of $\mu_{ij}$ is at most $\sum_g\omega_g^2\sum_bw_b^2/4$; with $J=6$ and $B=25$ the standard error is at most $0.042$, and that of $\mu_{ij}-\mu^{\mathrm{core}}_{ij}$ at most $0.060$.

\begin{proposition}\label{prop:comemb}
(i) $\mu$ is a reflexive and symmetric fuzzy relation with values in $[0,1]$. (ii) If all ties are crisp, $\mu$ and $\mu^{\mathrm{core}}$ have the same distribution; if moreover $\Cl$ is deterministic, $\mu_{ij}=\mathbb 1\{i\sim_{\Cl(\Phi(M,M^\top))}j\}$. (iii) If $\Cl$ is permutation equivariant in distribution, relabeling the actors permutes the rows and columns of $\mu$ in distribution, and $\mathrm{E}(\mu)$ is permuted exactly. (iv) $\mu$ is positive semidefinite, and $d_{ij}=\sqrt{2(1-\mu_{ij})}$ is a Euclidean (pseudo)metric on $V$.
\end{proposition}

The proof is in Appendix~A.

\paragraph{Consensus partition, membership and ambiguity.} Let $h_{(1)}\le\dots\le h_{(n-1)}$ be the merge heights of the average-linkage dendrogram of $1-\mu$. The lifetime of the partition into $K$ groups is $\ell_K=h_{(n-K+1)}-h_{(n-K)}$, and the consensus partition $\hat P=\{G_1,\dots,G_{\hat K}\}$ is the cut into $\hat K=\arg\max_{K\in\{2,\dots,8\}}\ell_K$ groups.\cite{fred2005} The criterion cannot return $\hat K=1$, so the absence of community structure cannot be detected through $\hat K$. By Proposition~\ref{prop:comemb}(iv), average linkage on $1-\mu$ operates on half the squared Euclidean distances $d_{ij}^2$. The membership degree of actor $i$ to group $k$ is $u_{ik}=|G_k\setminus\{i\}|^{-1}\sum_{j\in G_k\setminus\{i\}}\mu_{ij}$ (equal to $1$ when $i$ is the only member of $G_k$), normalized as $\tilde u_{ik}=u_{ik}/\sum_hu_{ih}$; it is an affinity index with respect to groups obtained a posteriori, not a posterior probability. The assignment ambiguity index $\eta_i=\frac2{n-1}\sum_{j\ne i}\min\{\mu_{ij},1-\mu_{ij}\}$ is the linear index of fuzziness\cite{kaufmann1975} of the fuzzy set $\mu_{i\cdot}$: it is $0$ when all co-memberships of $i$ are crisp and $1$ when they all equal $1/2$. It refers to the assignment of an actor, not to the vagueness of a fuzzy number.

\begin{algorithm}[t]
\caption{Fuzzy co-membership with modularity (FCo-L)}\label{alg:fco}
\begin{algorithmic}[1]
\Require $L,M,R$; $\Phi$; grid $\alpha_1,\dots,\alpha_J$; $B$; $\lambda$; algorithm $\Cl$
\For{$g=1,\dots,J$}
  \State $W^L\gets L+\alpha_g(M-L)$, $W^U\gets R-\alpha_g(R-M)$
  \For{$b=1,\dots,B$} \State build $W^{(b)}$ (Definition~\ref{def:comemb}); $P^{(b)}\gets\Cl(\Phi(W^{(b)},W^{(b)\top}))$ \EndFor
  \State $\pi(\alpha_g)\gets$ Eq.~(\ref{eq:mu}) with weights $w_b$
\EndFor
\State $\mu\gets\sum_g\omega_g\pi(\alpha_g)$; $\mu_{ii}\gets1$
\State $\hat K\gets\arg\max_K\ell_K$ on the average-linkage dendrogram of $1-\mu$; $\hat P\gets$ cut into $\hat K$ groups
\State \Return $\mu$, $\{\pi(\alpha_g)\}$, $\hat P$, $\tilde u$, $\eta$
\end{algorithmic}
\end{algorithm}

The procedure requires $JB$ Louvain runs, each nearly linear in the number of ties, and one average-linkage clustering; for networks of the size of social surveys it takes seconds. Any community detection algorithm can replace Louvain, for instance Leiden\cite{traag2019} when connected communities are required.

\section{Simulation Study}\label{sec:sim}

\subsection{Design}\label{sec:simdesign}

\begin{figure}[tbp]
\centerline{\includegraphics[width=0.8\textwidth]{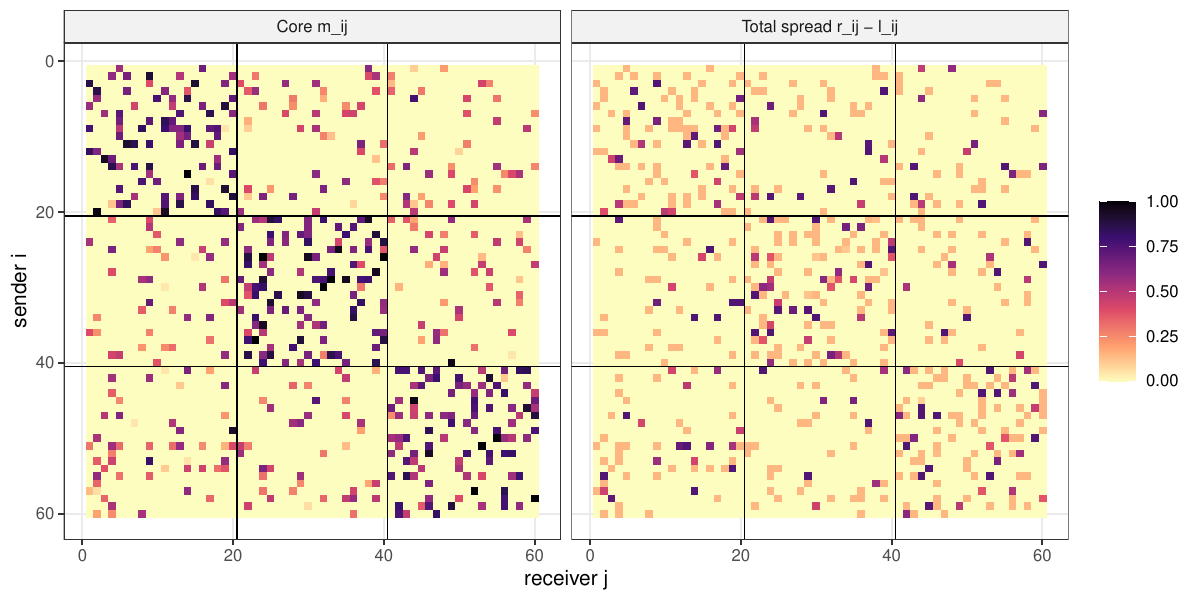}}
\caption{One simulated network of the main design ($n=60$, $p_{\mathrm{out}}=0.10$, informative spreads), with the actors ordered by true block. Left: cores $m_{ij}$ of the directed ties; the three diagonal blocks are denser and stronger than the off-diagonal ones. Right: total spreads $r_{ij}-l_{ij}$; wide supports correspond to the low-reliability ties and are scattered over the whole matrix, so that vagueness is unrelated to the position of the actors.}\label{fig:simnet}
\alttext{Two heat maps of a simulated fuzzy network, showing the cores and the total spreads of the ties with actors ordered by block.}
\end{figure}

Networks are generated from a directed stochastic block model\cite{holland1983} with $K=3$ balanced blocks and TFN ties. A directed tie exists with probability $0.25$ within blocks and $p_{\mathrm{out}}\in\{0.05,0.10\}$ between blocks, with latent strength $\mathcal U(0.50,0.90)$ within and $\mathcal U(0.20,0.60)$ between blocks. With probability $0.30$ a tie has low reliability (noise standard deviation $0.25$, otherwise $0.05$); the core is the latent strength plus noise, truncated to $[0.05,1]$, and the spread is $1.5$ times the noise standard deviation on both sides. In the \emph{informative} setting vague ties are the unreliable ones; in the \emph{uninformative} setting the spreads are permuted across ties. Crossing $n\in\{30,60\}$, $p_{\mathrm{out}}$ and the setting gives eight scenarios with $100$ replicates each; Fig.~\ref{fig:simnet} shows a simulated network. FCo-L uses $\alpha\in\{0,0.2,\dots,1\}$, $B=25$ and $\lambda=3/25$.

Four further designs have $100$ replicates each. \emph{Null scenarios}: ties without blocks, for $n\in\{30,60\}$, with probability $0.15$ (the density of the scenario $n=30$, $p_{\mathrm{out}}=0.10$), latent strength $\mathcal U(0.20,0.90)$ and the reliability and spread mechanism of the informative setting. \emph{Sensitivity}: on the networks of the two scenarios with $n=30$ and $p_{\mathrm{out}}=0.10$, FCo-L is recomputed with uniform realizations only, with $B=50$ (same $\lambda$, which changes only the Monte Carlo error), with the grid $\alpha\in\{0,0.1,\dots,1\}$ and with the upper bound of $\hat K$ raised from $8$ to $15$. \emph{Boundary actors}: in a block model with $n\in\{30,60\}$ and $p_{\mathrm{out}}=0.05$, $10\%$ of the actors ($3$ and $6$) have a home block, which is their true label, and a second block; their ties with both blocks exist with probability $0.15$ and have latent strength $\mathcal U(0.35,0.75)$. All ties of a set of vague actors then receive the large spread $1.5\times0.25$; the vague actors are the boundary actors (\emph{aligned} arm, favorable to FCo-L by construction) or as many other actors (\emph{misaligned} arm), with the same cores in the two arms. \emph{Asymmetric TFNs}: the TFNs above are symmetric except near $0$ and $1$, whereas the real ratings are markedly asymmetric (Section~\ref{sec:real}). With the block model of the main study, the support of each tie spans the interval between its core $m_{ij}$ and its latent strength $w^*_{ij}$, widened by $0.025$ on both sides and clipped to $[0,1]$; about $80\%$ of the ties then have $|(r_{ij}-m_{ij})-(m_{ij}-l_{ij})|>(r_{ij}-l_{ij})/4$, close to the $84\%$ and $94\%$ of the real networks. In the \emph{informative} arm the longer side of each TFN points towards the latent strength; in the \emph{random-direction} arm left and right spreads are swapped with probability $1/2$, which preserves widths and degree of asymmetry. Since Perturb-tie matches the FCo-L perturbation tie by tie but is symmetric, comparing the two arms isolates the contribution of the direction of the asymmetry.

\subsection{Competing methods and criteria}\label{sec:competitors}

FCo-L is compared with Consensus-core, Perturb-core and Perturb-tie (Definition~\ref{def:control}), with the same consensus step; with Louvain-core and Louvain-CoG, single runs on the projected cores and centers of gravity $(l+m+r)/3$; and with IW-Louvain,\cite{alves2024} applied to the supports after the projection $\Phi=\max$, with interval arithmetic in both phases and the expected interval weights of their Definition 4.6.

Recovery is measured by the adjusted Rand index (ARI)\cite{hubert1985} and normalized mutual information\cite{strehl2002} with respect to the true blocks, and by the proportion of replicates with three communities. The co-membership relation is evaluated over pairs by the Brier score and the area under the ROC curve (AUC) of $\mu_{ij}$ for the indicator of the same block. The ambiguity index is evaluated against a target common to all methods and defined by the data-generating process: the share of the latent strength of the projected ties of actor $i$ that goes outside its true block. We report the Spearman correlation between $\eta_i$ and this share (\emph{structural Spearman}); in the boundary-actor design, the AUC of $\eta_i$ for the boundary actors. We also report the AUC of $\eta_i$ for the actors misclassified by the same method, which is specific to each relation. Paired differences on the same networks are reported with Monte Carlo standard errors and $z$ ratios; for a criterion compared in eight scenarios, or in four scenarios and two arms, the two-sided Bonferroni threshold at level $0.05$ is $|z|>2.73$, and for the $48$ comparisons of ARI between FCo-L and all competitors it is $|z|>3.28$.

\subsection{Results}\label{sec:simresults}

\begin{figure}[tbp]
\centerline{\includegraphics[width=0.95\textwidth]{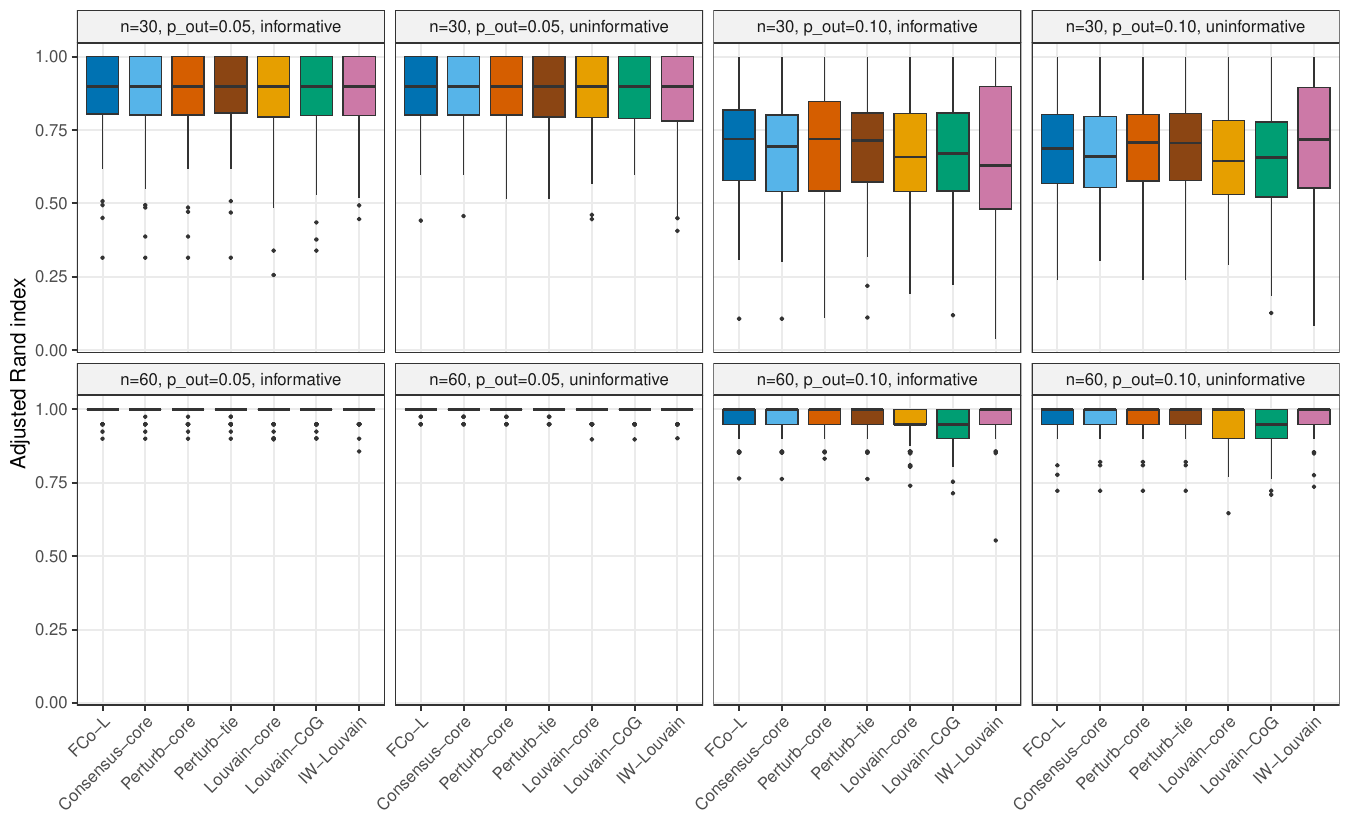}}
\caption{Main simulation design: distribution of the adjusted Rand index between the partition of each method and the true blocks over $100$ replicates, for the eight scenarios defined by network size $n$, between-block tie probability $p_{\mathrm{out}}$ and informative or uninformative spreads. The four consensus methods (FCo-L, Consensus-core, Perturb-core, Perturb-tie) are compared with single Louvain runs on the cores (Louvain-core) and on the centers of gravity (Louvain-CoG) and with the interval-weighted Louvain algorithm (IW-Louvain).}\label{fig:ari}
\alttext{Box plots of the adjusted Rand index of seven methods in eight simulation scenarios.}
\end{figure}

\begin{figure}[tbp]
\centerline{\includegraphics[width=0.95\textwidth]{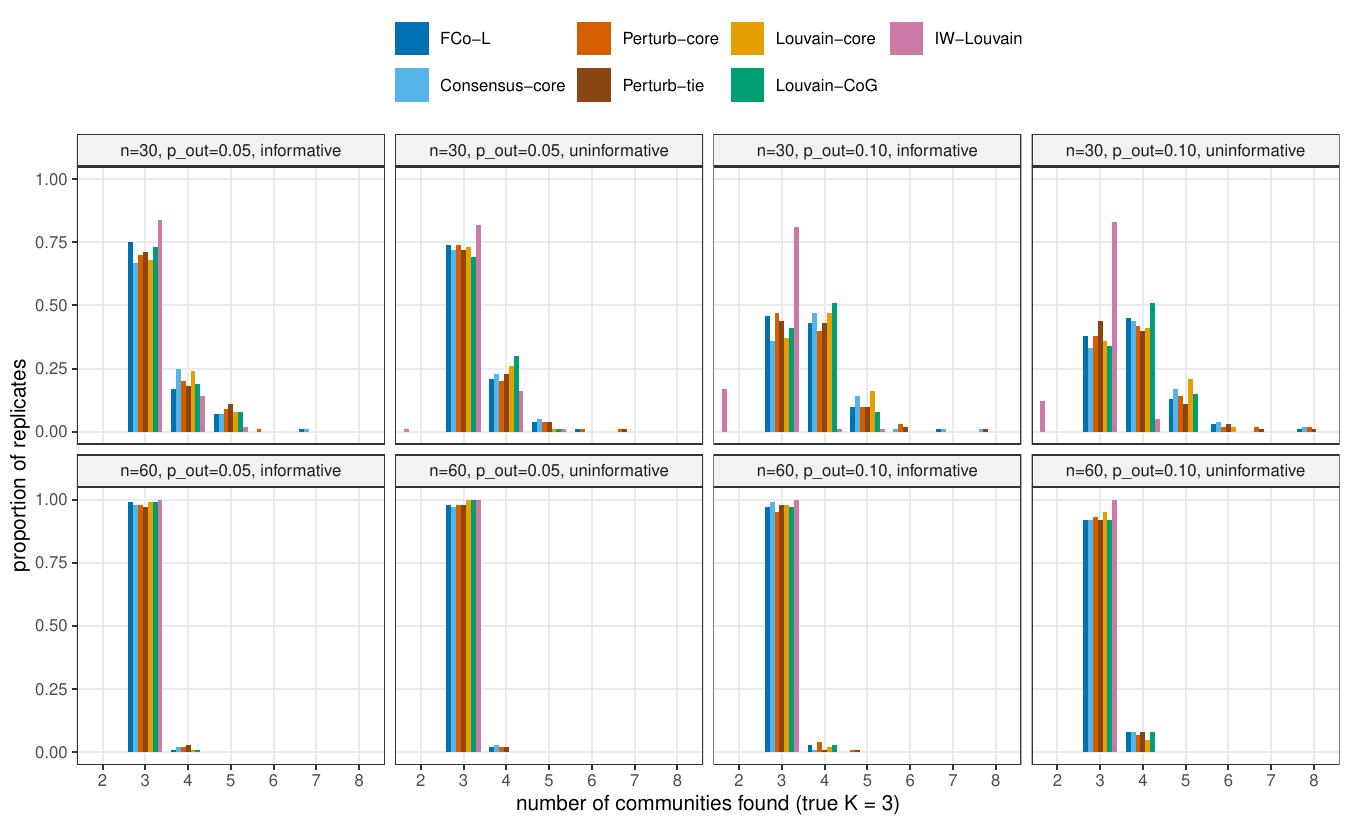}}
\caption{Main simulation design: proportion of the $100$ replicates in which each method returns a given number of communities; the true number is $K=3$. With $n=60$ all methods concentrate on three communities; with $n=30$ and $p_{\mathrm{out}}=0.10$ IW-Louvain returns three communities most often, while the other methods frequently return four or five, splitting a block.}\label{fig:ncomm}
\alttext{Bar charts of the distribution of the number of communities found by seven methods in eight simulation scenarios.}
\end{figure}

\paragraph{Recovery.} With $n=60$ all methods are close to perfect recovery when $p_{\mathrm{out}}=0.05$ (mean ARI from $0.992$ to $0.997$) and have mean ARI of at least $0.945$ when $p_{\mathrm{out}}=0.10$ (Fig.~\ref{fig:ari}, Table~\ref{tab:rec}). With $n=30$ and $p_{\mathrm{out}}=0.10$ the mean ARI is between $0.643$ and $0.698$ for every method. FCo-L improves on single Louvain runs on the cores in all scenarios, by $0.002$ to $0.032$ ($z>3.28$ in two scenarios), and differs from IW-Louvain by $-0.002$ to $0.030$ ($|z|\le2.64$; Table~\ref{tab:paired}). Against Consensus-core the paired difference is $0.027$ ($z=3.84$) with $n=30$, $p_{\mathrm{out}}=0.10$ and informative spreads, $0.013$ ($z=1.81$) with uninformative spreads, and at most $0.008$ elsewhere (Table~\ref{tab:core}). Against Perturb-core and Perturb-tie the differences are between $-0.010$ and $0.006$ ($|z|\le1.49$) and between $-0.012$ and $0.007$ ($|z|\le2.47$) (Tables~\ref{tab:pert} and~\ref{tab:tie}): the improvement over the consensus of the cores is obtained equally by perturbing the cores by the same amount. With $n=60$ all methods return three communities in at least $92\%$ of the replicates. With $n=30$ and $p_{\mathrm{out}}=0.10$ IW-Louvain does so in $81\%$ and $83\%$ of the replicates, FCo-L in $46\%$ and $38\%$; the other methods tend to split a block and IW-Louvain to merge two (Fig.~\ref{fig:ncomm}).

\begin{table}[tbp]
\tbl{Main simulation design, $100$ replicates per scenario. Recovery of the true blocks: adjusted Rand index (ARI) and normalized mutual information (NMI), mean (Monte Carlo standard error); proportion of replicates in which the method returns $\hat K=3$ communities and mean $\hat K$. Co-membership: Brier score, mean (standard error), and pairwise AUC of the relation for the indicator of the same block. Ambiguity: AUC of $\eta_i$ for each relation's own misclassified actors, with the number of replicates in which it is defined in brackets, and structural Spearman correlation between $\eta_i$ and the share of the latent strength of actor $i$ that goes outside its true block, mean (standard error); these criteria are not defined for single partitions.\label{tab:rec}}
{\adjustbox{max width=\textwidth,max totalheight=0.85\textheight}{\input{tables/Table_01_Simulation_Recovery_CoMembership.tex}}}
\end{table}

\begin{table}[tbp]
\tbl{Main simulation design: paired differences in ARI between FCo-L and each competitor on the same networks, mean (Monte Carlo standard error) over $100$ replicates. Positive values favor FCo-L; with $48$ comparisons the Bonferroni threshold for the ratio of the mean to its standard error is $3.28$.\label{tab:paired}}
{\adjustbox{max width=\textwidth,max totalheight=0.85\textheight}{\input{tables/Table_02_Paired_Differences.tex}}}
\end{table}

\begin{figure}[tbp]
\centerline{\includegraphics[width=0.95\textwidth]{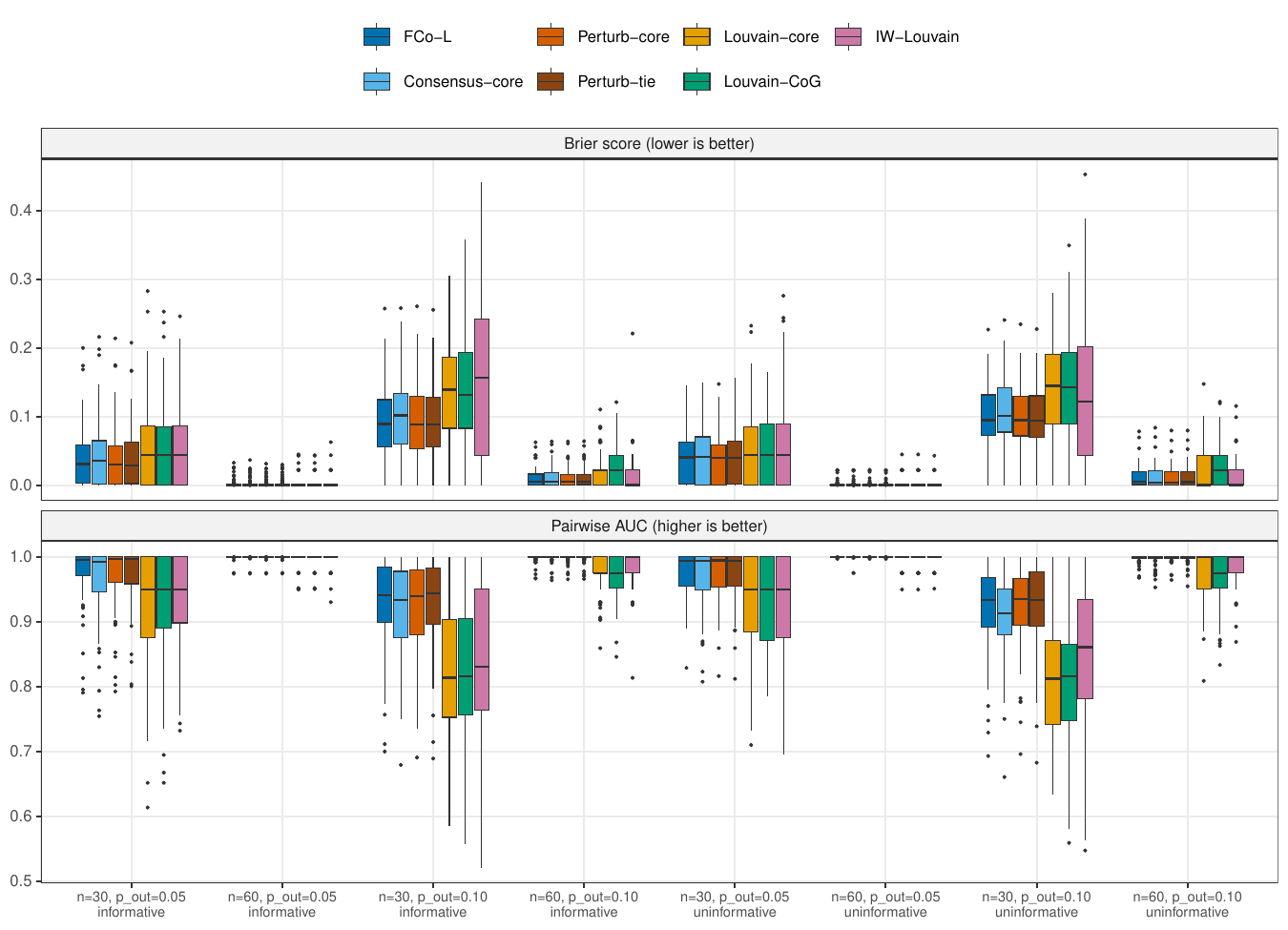}}
\caption{Main simulation design: quality of the co-membership relations as predictors of the indicator that two actors belong to the same true block. Top: Brier score (lower is better); bottom: pairwise AUC (higher is better). For single partitions the relation is the crisp co-membership matrix; for the consensus methods it is the fuzzy relation $\mu$ or its analogue.}\label{fig:comemb}
\alttext{Box plots of the Brier score and pairwise AUC of the co-membership relations in eight simulation scenarios.}
\end{figure}

\begin{figure}[tbp]
\centerline{\includegraphics[width=0.8\textwidth]{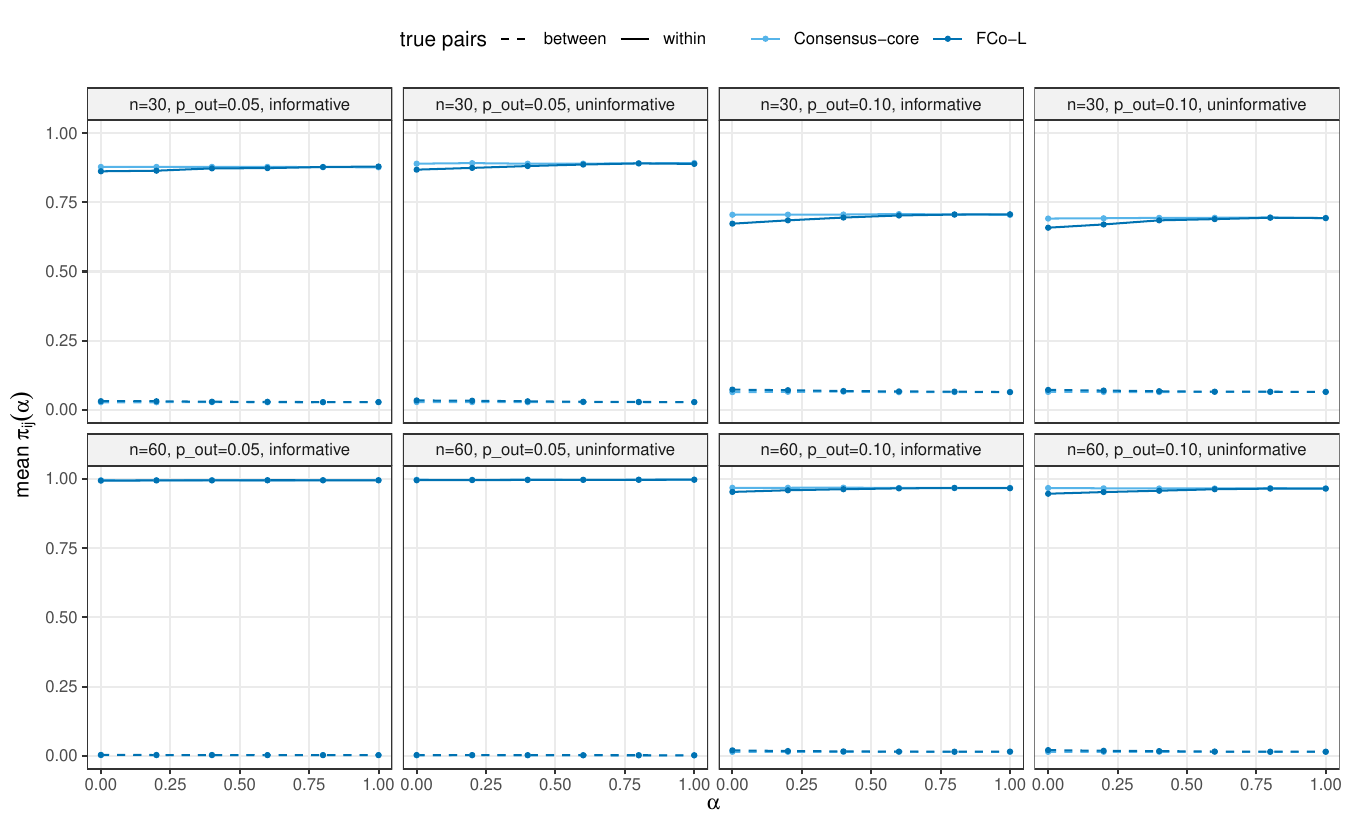}}
\caption{Main simulation design: average level-wise co-membership $\pi_{ij}(\alpha)$ over pairs of actors in the same block (solid lines) and in different blocks (dashed lines), for FCo-L and Consensus-core. Consensus-core does not depend on $\alpha$ and FCo-L coincides with it at $\alpha=1$, where every realization equals the cores; lower levels allow wider realizations and lower the co-membership of pairs in the same block.}\label{fig:alpha}
\alttext{Line plots of the average level-wise co-membership as a function of alpha for same-block and different-block pairs.}
\end{figure}

\paragraph{Co-membership.} The fuzzy relations have a lower Brier score than the crisp co-membership of single partitions (Fig.~\ref{fig:comemb}, Table~\ref{tab:rec}): with $n=30$ and $p_{\mathrm{out}}=0.10$ it is $0.094$ and $0.099$ for FCo-L and Perturb-core, $0.093$ and $0.097$ for Perturb-tie, $0.102$ and $0.108$ for Consensus-core and between $0.133$ and $0.152$ for single partitions. The reduction of FCo-L over Consensus-core is consistent for $n=30$ (from $-0.003$ to $-0.009$, $|z|$ from $3.26$ to $7.76$) and vanishes for $n=60$. Against Perturb-core the differences are at most $0.0011$ in absolute value; with uninformative spreads and $p_{\mathrm{out}}=0.10$ Perturb-tie has a lower Brier score than FCo-L ($0.0021$, $z=3.18$, for $n=30$; $0.0005$, $z=2.91$, for $n=60$). For pairs in the same block the level-wise co-membership of FCo-L decreases as $\alpha$ decreases, by about $0.03$ at $\alpha=0$ with $n=30$ and $p_{\mathrm{out}}=0.10$ (Fig.~\ref{fig:alpha}).

\begin{figure}[tbp]
\centerline{\includegraphics[width=0.85\textwidth]{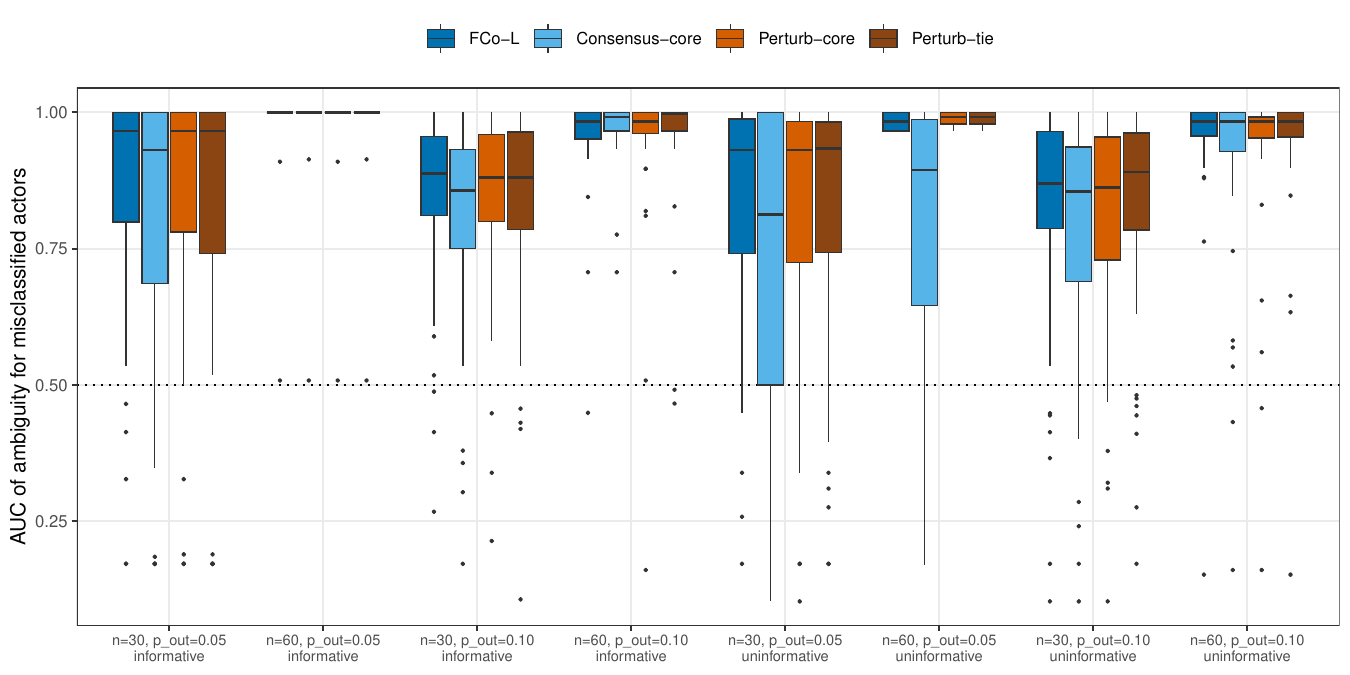}}
\caption{Main simulation design: AUC of the ambiguity index $\eta_i$ of the four consensus relations as a classifier of the actors misclassified by the consensus partition of the same relation, over the replicates in which the AUC is defined; the dotted line marks $0.5$. The set of misclassified actors differs across relations, so the panels describe how well each relation signals its own errors and do not compare the methods on a common outcome.}\label{fig:ambmis}
\alttext{Box plots of the AUC of the ambiguity index for misclassified actors for four consensus relations in eight scenarios.}
\end{figure}

\paragraph{Ambiguity.} The structural Spearman correlation of FCo-L is between $0.31$ and $0.44$ for $n=30$ and between $0.09$ and $0.34$ for $n=60$ (Table~\ref{tab:rec}). It exceeds that of Consensus-core in all scenarios (by $0.029$ to $0.141$, $z>2.73$ in six), that of Perturb-core in all scenarios (by $0.014$ to $0.067$, $z>2.73$ in five) and that of Perturb-tie in all scenarios, by smaller amounts (up to $0.041$, $z>2.73$ in four). The TFNs of this design are symmetric, so the advantage over Perturb-tie derives from the mixture of perturbation sizes over $\alpha$ and from the extreme realizations, not from asymmetry. Informative and uninformative spreads give results of the same size for all criteria: the low-reliability ties are assigned independently of the actors, and the spreads carry no information on which actors are difficult to assign. The index of FCo-L also signals its own misclassified actors (AUC $0.836$--$0.857$ for $n=30$, above $0.94$ for $n=60$, where they are rare; Table~\ref{tab:rec}, Fig.~\ref{fig:ambmis}).

\begin{figure}[tbp]
\centerline{\includegraphics[width=0.85\textwidth]{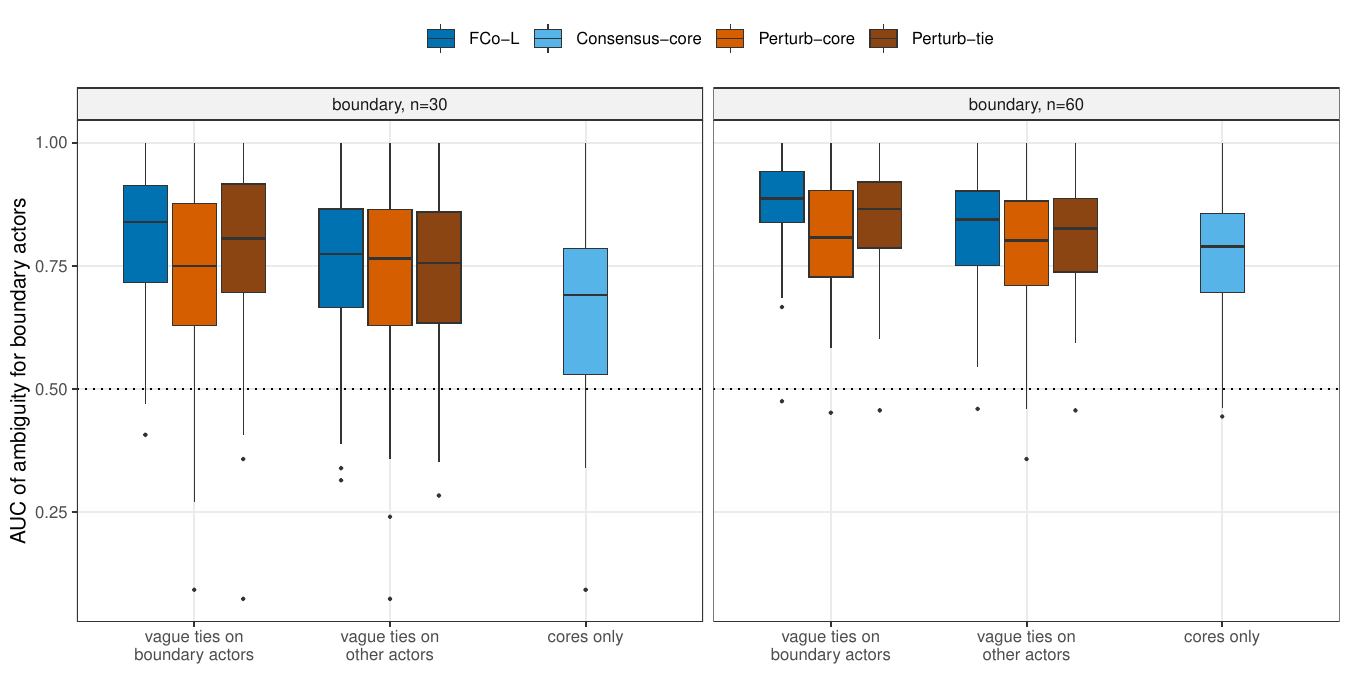}}
\caption{Boundary-actor design: AUC of the ambiguity index $\eta_i$ as a classifier of the boundary actors, $100$ replicates for $n=30$ (left) and $n=60$ (right). Within each panel, the first group corresponds to the aligned arm, in which the boundary actors also have the vague ties, the second to the misaligned arm, in which the vague ties belong to as many other actors and the cores are unchanged; Consensus-core uses only the cores and is the same in the two arms. The dotted line marks $0.5$.}\label{fig:bnd}
\alttext{Box plots of the AUC of the ambiguity index for boundary actors under aligned and misaligned vagueness.}
\end{figure}

\paragraph{Boundary actors.} Boundary actors are misclassified in $46$--$52\%$ of the cases, the other actors in at most $5.4\%$ (Table~\ref{tab:bnd}), and no method improves the partition ($|z|\le2.55$; Table~\ref{tab:bndpaired}). With the vagueness on the boundary actors, the AUC of $\eta_i$ for boundary actors is $0.807$ and $0.874$ for FCo-L ($n=30$ and $60$), $0.725$ and $0.811$ for Perturb-core, $0.783$ and $0.855$ for Perturb-tie and $0.670$ and $0.774$ for Consensus-core (Fig.~\ref{fig:bnd}); the differences of FCo-L are $0.082$ ($z=7.91$) and $0.063$ ($z=7.24$) against Perturb-core and $0.025$ ($z=3.48$) and $0.019$ ($z=3.01$) against Perturb-tie. Most of the gain over the homogeneous perturbation is thus obtained by a tie-specific scale, the remainder by the shape of the exploration. With misplaced vagueness the AUC of FCo-L is $0.749$ and $0.826$; it remains above Consensus-core, but the differences with Perturb-tie ($0.012$, $z\le2.00$) are not distinguishable from zero.

\begin{table}[tbp]
\tbl{Boundary-actor design, $100$ replicates for each $n$: ARI; AUC of the ambiguity index for boundary actors and for misclassified actors; mean ambiguity and share of misclassified actors among boundary and other actors. \emph{Aligned}: the vague ties are those of the boundary actors; \emph{misaligned}: those of as many other actors; \emph{common}: methods that use only the cores, whose results do not depend on the arm.\label{tab:bnd}}
{\adjustbox{max width=\textwidth,max totalheight=0.85\textheight}{\input{tables/Table_03_Boundary_Summary.tex}}}
\end{table}

\begin{table}[tbp]
\tbl{Boundary-actor design: paired differences between FCo-L and each consensus benchmark on the same networks, for the ARI and for the AUC of the ambiguity index for boundary actors, mean (standard error) and $z$ ratio. Positive values favor FCo-L; the Bonferroni threshold for eight comparisons is $|z|>2.73$.\label{tab:bndpaired}}
{\adjustbox{max width=\textwidth,max totalheight=0.85\textheight}{\input{tables/Table_04_Boundary_Paired.tex}}}
\end{table}

\begin{figure}[tbp]
\centerline{\includegraphics[width=0.95\textwidth]{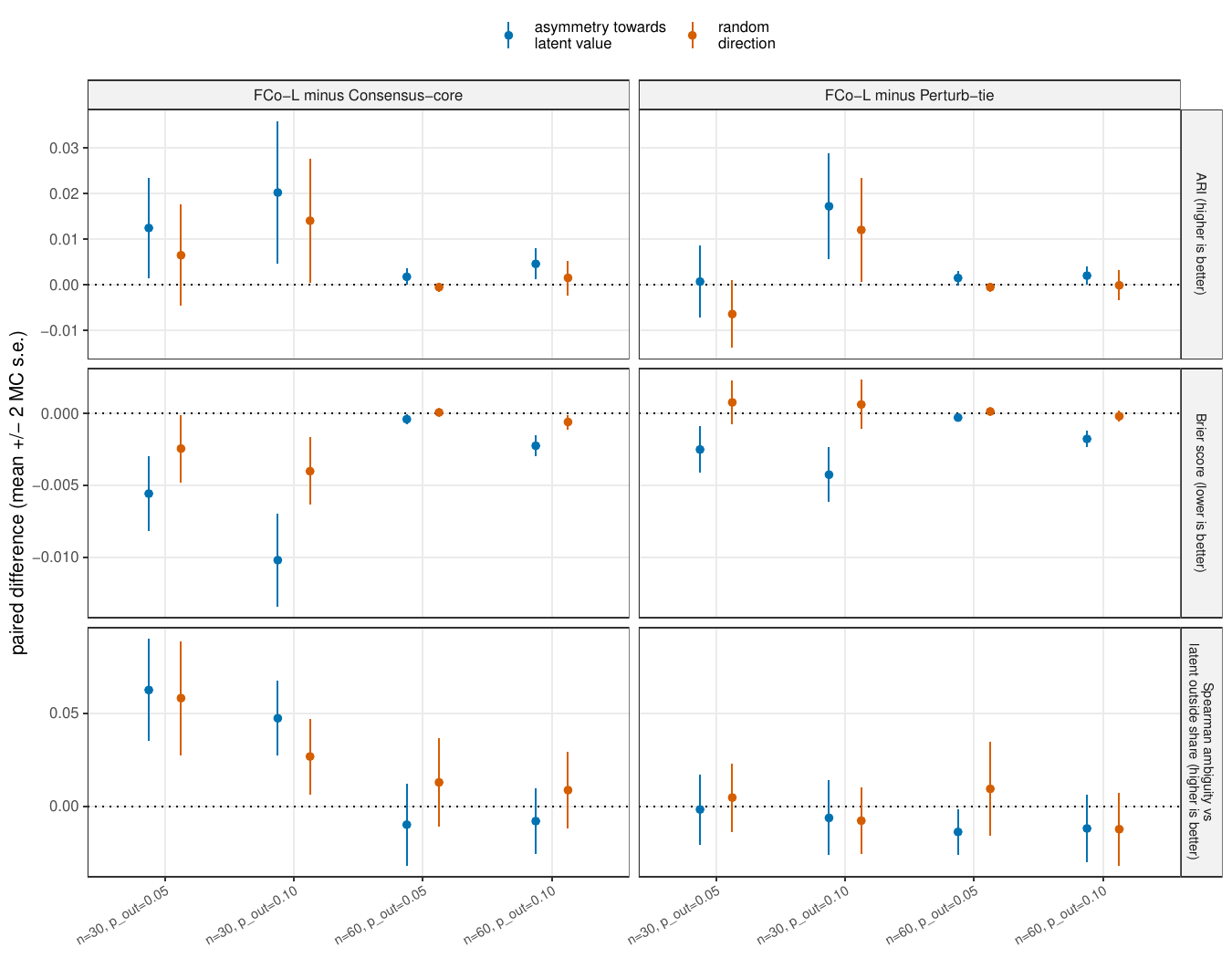}}
\caption{Asymmetric-TFN design: paired differences of FCo-L against Consensus-core (left) and Perturb-tie (right) for ARI, Brier score and structural Spearman correlation, mean $\pm2$ Monte Carlo standard errors over $100$ replicates. Blue: informative arm, in which the longer side of each TFN points towards the latent strength; orange: random-direction arm, with the same widths and degree of asymmetry. Negative values in the Brier panels and positive values in the other panels favor FCo-L.}\label{fig:asym}
\alttext{Paired differences of FCo-L against Consensus-core and Perturb-tie for ARI, Brier score and structural Spearman correlation in the asymmetric design.}
\end{figure}

\paragraph{Asymmetric TFNs.} The direction of the asymmetry affects the Brier score of FCo-L and not that of the benchmarks (Fig.~\ref{fig:asym}, Tables~\ref{tab:asym} and~\ref{tab:asympaired}). When the longer side points towards the latent strength, FCo-L has a lower Brier score than Perturb-tie in all four scenarios: $-0.0025$ ($z=-3.09$) and $-0.0043$ ($z=-4.48$) for $n=30$ with $p_{\mathrm{out}}=0.05$ and $0.10$, $-0.0003$ ($z=-2.07$) and $-0.0018$ ($z=-6.48$) for $n=60$, reductions of about $5\%$ for $n=30$ and of $17$--$21\%$ of very small Brier scores for $n=60$. With random direction the differences are between $-0.0002$ and $0.0008$ ($|z|\le1.37$). Between the two arms, the Brier score of FCo-L decreases in all scenarios ($z$ from $-2.78$ to $-5.42$), that of Perturb-tie does not change ($|z|\le2.03$). The partition and the ambiguity index do not benefit: the differences in ARI with Perturb-tie are at most $0.017$ ($z\le2.97$) and those in structural Spearman correlation are between $-0.014$ and $-0.002$. IW-Louvain uses the asymmetric supports more effectively for the partition: in the informative arm with $n=30$ its mean ARI is $0.897$ and $0.754$, against $0.875$ and $0.716$ for FCo-L ($z=-2.85$ and $-2.67$), and it returns three communities more often.

\paragraph{Sensitivity and null scenarios.} Across the sensitivity variants the mean ARI of FCo-L varies by at most $0.006$, the Brier score by at most $0.0005$ (Table~\ref{tab:sens}). Without communities all methods except IW-Louvain return several groups (FCo-L on average $4.53$ and $5.44$; Table~\ref{tab:null}), and the mean co-membership over pairs is $0.22$ and $0.19$ for every method, including single partitions, so it reflects the number of groups returned rather than the absence of structure. The mean ambiguity of FCo-L ($0.258$ and $0.307$) exceeds that of the block models with the same $n$ ($0.090$ and $0.018$), but Consensus-core shows a similar increase. Without a reference distribution the relation describes the instability of the groups but does not establish that a network has no community structure.

\begin{table}[tbp]
\tbl{Sensitivity of FCo-L in the two scenarios with $n=30$ and $p_{\mathrm{out}}=0.10$, same $100$ networks for all variants. The default uses mixed realizations, $B=25$, $\alpha$ step $0.2$ and $\hat K\le8$; each variant changes one setting. ARI, mean (standard error); proportion of replicates with three communities and mean number of communities; Brier score; AUC of the ambiguity index for misclassified actors, with the number of replicates in which it is defined in brackets.\label{tab:sens}}
{\adjustbox{max width=\textwidth,max totalheight=0.85\textheight}{\input{tables/Table_05_Sensitivity.tex}}}
\end{table}

\begin{table}[tbp]
\tbl{Null scenarios, networks without communities, $100$ replicates for each $n$: proportion of replicates with two groups, mean number of groups, mean co-membership over pairs (crisp co-membership for single partitions) and mean ambiguity; the last column reports, for comparison, the mean ambiguity of the same method in the four block-model scenarios of the main design with the same $n$.\label{tab:null}}
{\adjustbox{max width=\textwidth,max totalheight=0.85\textheight}{\input{tables/Table_06_Null_Scenario.tex}}}
\end{table}

\section{Application to Departmental Collaboration Networks}\label{sec:real}

\subsection{Data and analysis}\label{sec:data}

The data were collected with an interactive web questionnaire among the professors of two departments of the University ``G.\ d'Annunzio'' of Chieti-Pescara, Italy: the Department of Economic Studies (A), whose network was analyzed by Porreca et al.,\cite{porreca2025} and the Department of Medical, Oral and Biotechnological Sciences (B). Following the Dynamic Fuzzy Rating Tracker,\cite{calcagni2014} each respondent rated the collaboration with each colleague by moving the mouse on a semicircular five-point scale ($1$ never, $2$ seldom, $3$ occasionally, $4$ often, $5$ constantly), and the pointer trajectory was recorded until the final click. Each rating is the TFN (minimum, final position, maximum) of the trajectory on $[1,5]$, mapped to $[0,1]$ by $(x-1)/4$. A tie is observed when its support is not $\{0\}$; later ratings of the same colleague replace earlier ones, and actors with no observed ties are excluded. The networks have $51$ actors and $219$ directed ties (A) and $55$ actors and $254$ ties (B); $24$ and $28$ actors responded. Pairs without a rating are coded as null ties, which treats the absence of a reported collaboration as absence of collaboration; this is not equivalent when a colleague is not rated because of non-response.

We apply FCo-L and the benchmarks with the settings of the simulation and repeat the consensus relations $20$ times with independent seeds to assess the Monte Carlo error. As a reference for the ambiguity index, FCo-L is recomputed on $20$ networks in which the ratings of each respondent are randomly permuted among the other actors, which preserves the number and values of the ratings and destroys the group structure. To assess the dependence on non-response, FCo-L is recomputed (i) with the ties between pairs of non-respondents unknown, i.e.\ vacuous possibility distributions with $\alpha$-cuts $[0,1]$, (ii) with all ties that non-respondents would have sent unknown, and (iii) on the subnetwork of the respondents, where every tie is observed in both directions. We also use the projection $\Phi=$ mean and the participation coefficient $\mathrm{PC}_i=1-\sum_k(s_{ik}/s_i)^2$,\cite{guimera2005} where $s_{ik}$ is the strength of $i$ towards group $k$.

\subsection{Results}\label{sec:realresults}

\begin{figure}[tbp]
\centerline{\includegraphics[width=0.75\textwidth]{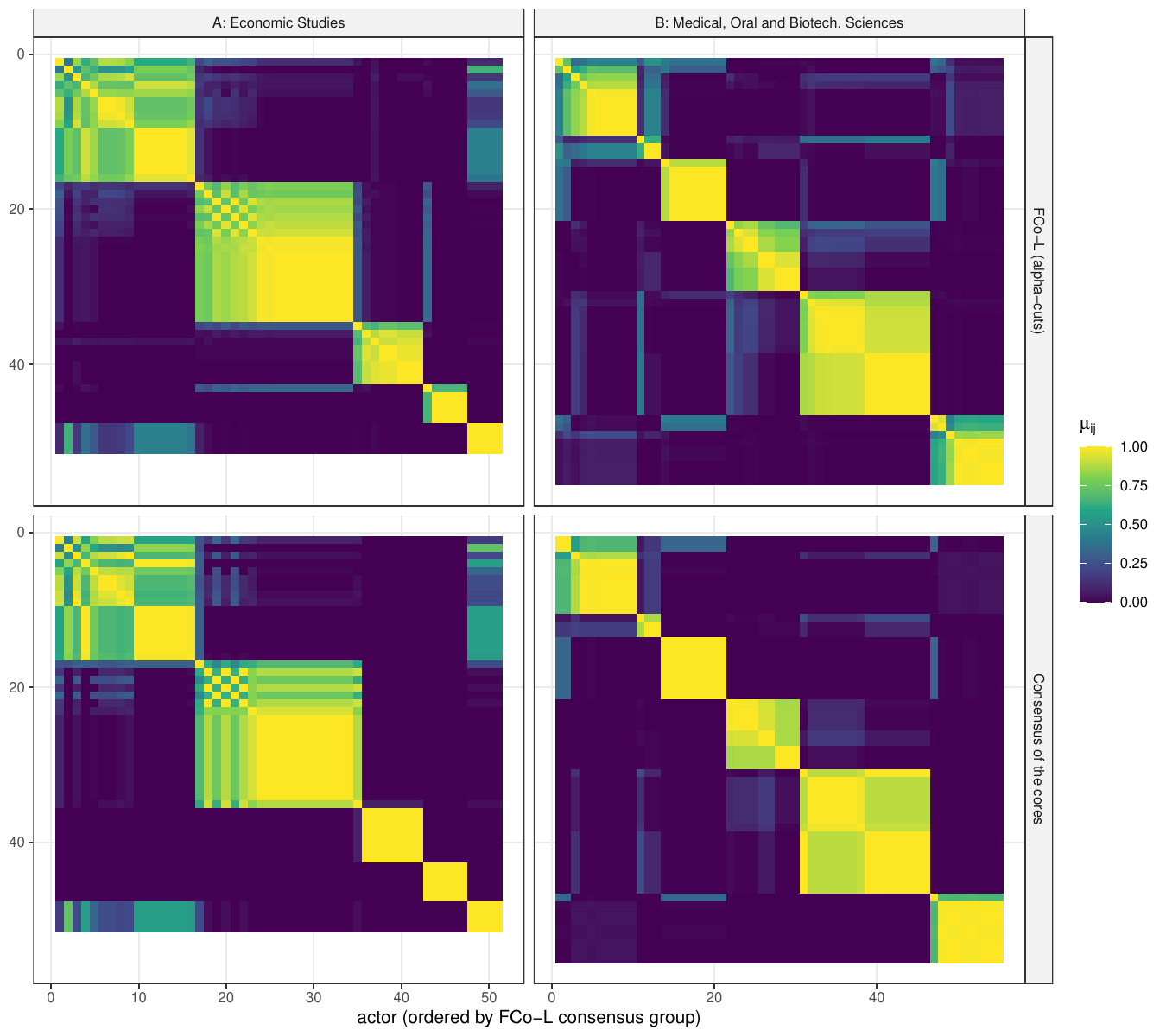}}
\caption{Real networks: fuzzy co-membership relations $\mu$ of FCo-L (top) and of Consensus-core (bottom) in departments A and B, with the actors ordered by FCo-L consensus group. Diagonal blocks close to one correspond to stable groups; off-diagonal cells with intermediate values identify actors whose assignment changes across realizations, which are more frequent under FCo-L.}\label{fig:rdmu}
\alttext{Heat maps of the fuzzy co-membership relations of FCo-L and Consensus-core in two departments.}
\end{figure}

\begin{figure}[tbp]
\centerline{\includegraphics[width=0.95\textwidth]{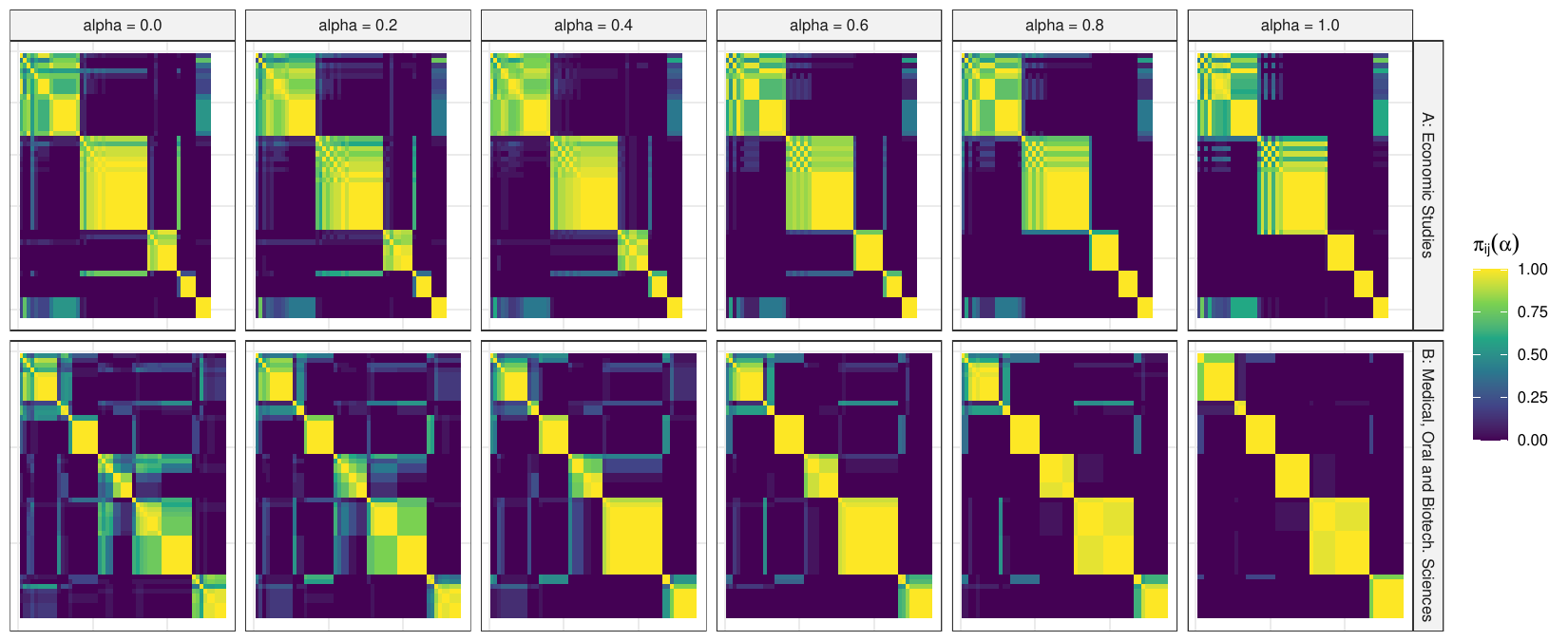}}
\caption{Real networks: level-wise co-membership $\pi_{ij}(\alpha)$ of FCo-L for $\alpha=0,0.2,\dots,1$ in departments A (top) and B (bottom), actors ordered by FCo-L consensus group. At $\alpha=1$ all realizations equal the cores and only the randomness of the Louvain algorithm is present; lower levels add the vagueness of the ratings through wider $\alpha$-cuts. The relation $\mu$ of Fig.~\ref{fig:rdmu} is the integral of these matrices over $\alpha$.}\label{fig:rdalpha}
\alttext{Heat maps of the level-wise co-membership of FCo-L at six alpha levels in two departments.}
\end{figure}

The ratings are markedly asymmetric: in $94\%$ of the ties of A and $84\%$ of those of B the asymmetry exceeds a quarter of the support width, and on average the longer side lies below the core (Table~\ref{tab:rd}). The consensus partition of FCo-L has five groups in A and six in B (Fig.~\ref{fig:rdmu}). Its agreement with the partitions of the other methods ranges from ARI $0.616$ (Louvain-core) to $0.948$ (Perturb-core) in A and from $0.623$ (IW-Louvain) to $1$ (Consensus-core and Louvain-core) in B (Table~\ref{tab:rdagree}). In both departments the intermediate co-membership degrees between groups increase as $\alpha$ decreases, more markedly in B, where the groups are almost crisp at $\alpha=1$ (Fig.~\ref{fig:rdalpha}). The observed networks are less ambiguous than networks with permuted ratings (Table~\ref{tab:rdperm}): the mean ambiguity of FCo-L is $0.137$ in A and $0.119$ in B, against a minimum of $0.213$ and $0.212$ over the permutations, which nevertheless produce five or six groups on average. Over $20$ repetitions the standard deviation of $\mu_{ij}$ is at most $0.049$ and the mean ARI between partitions is $0.940$ and $0.961$, lower than for Consensus-core ($0.977$ and $1$) (Table~\ref{tab:rdstab}).

\begin{table}[tbp]
\tbl{Real networks: number of actors, respondents and directed ties; number of communities by method; group sizes of the FCo-L consensus partition; mean ambiguity of the four consensus relations; common perturbation size $\delta$ of Perturb-core; descriptive statistics of the TFN ratings on $[0,1]$, including the share of ties whose asymmetry $|(r-m)-(m-l)|$ exceeds a quarter of the support width.\label{tab:rd}}
{\adjustbox{max width=\textwidth,max totalheight=0.85\textheight}{\input{tables/Table_RealData_01_Summary.tex}}}
\end{table}

\begin{table}[tbp]
\tbl{Real networks: agreement between the FCo-L consensus partition and the partition of each other method in the main run (ARI and NMI), number of groups, and ARI with the FCo-L partition over $20$ independent repetitions of each method, mean [minimum, maximum], with the range of the number of groups over the repetitions.\label{tab:rdagree}}
{\adjustbox{max width=\textwidth,max totalheight=0.85\textheight}{\input{tables/Table_RealData_02_Agreement.tex}}}
\end{table}

\begin{table}[tbp]
\tbl{Real networks: FCo-L on the observed networks and under random permutation of the ratings of each respondent among the other actors, which preserves the number and values of the ratings and destroys the group structure; $20$ permutations. Mean ambiguity (mean [minimum, maximum] over permutations), number of groups and mean co-membership over pairs.\label{tab:rdperm}}
{\adjustbox{max width=\textwidth,max totalheight=0.85\textheight}{\input{tables/Table_RealData_03_Permutation_Reference.tex}}}
\end{table}

\begin{table}[tbp]
\tbl{Real networks: Monte Carlo stability of FCo-L and Consensus-core over $20$ independent repetitions; actors whose mean ambiguity over repetitions is higher or lower under FCo-L than under Consensus-core; sensitivity of FCo-L to the projection $\Phi$; Spearman correlation of the ambiguity index with the participation coefficient and between the repetition means of the ambiguity index of FCo-L and of the perturbation benchmarks.\label{tab:rdstab}}
{\adjustbox{max width=\textwidth,max totalheight=0.85\textheight}{\input{tables/Table_RealData_04_Stability.tex}}}
\end{table}

The partition depends on choices that precede the method. With $\Phi=$ mean, the ARI with the main partition is $0.640$ and $0.629$ (Table~\ref{tab:rdstab}). When the ties between pairs of non-respondents ($28\%$ and $24\%$ of the pairs) are unknown, FCo-L returns two and three groups unrelated to the main partition (ARI $-0.025$ and $0.057$): the vacuous ties join the non-respondents in a dense block, and the partition reflects the response status (Table~\ref{tab:rdunk}). On the subnetwork of the respondents FCo-L returns four groups, with ARI $0.933$ in A and $0.712$ in B against the main partition restricted to the same actors. The groups among the respondents of A are robust to the coding of non-response, those of B less so.

\begin{table}[tbp]
\tbl{Real networks, coding of non-response: FCo-L with null ties for unrated pairs (main analysis), with unknown ties (vacuous possibility distributions) between pairs of non-respondents, with unknown ties for all ties sent by non-respondents, and on the subnetwork of the respondents. Share of pairs with unknown ties, number of groups, ARI with the main partition (restricted to the respondents in the last row), mean ambiguity overall and among respondents and non-respondents.\label{tab:rdunk}}
{\adjustbox{max width=\textwidth,max totalheight=0.85\textheight}{\input{tables/Table_RealData_05_Unknown_Ties.tex}}}
\end{table}

\begin{figure}[tbp]
\centerline{\includegraphics[width=0.95\textwidth]{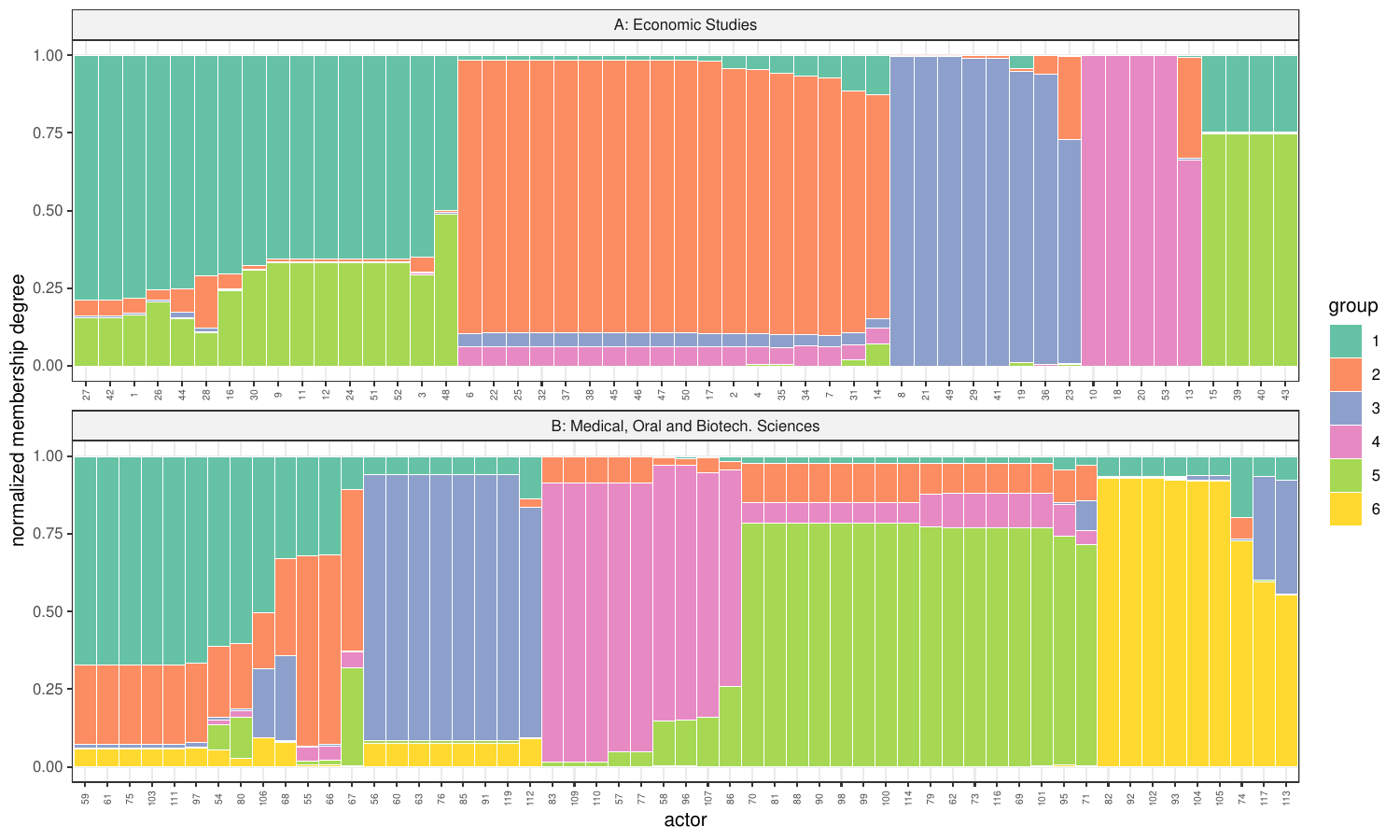}}
\caption{Real networks: normalized membership degrees $\tilde u_{ik}$ of each actor to the FCo-L consensus groups in departments A (top) and B (bottom), actors ordered by group. Each bar sums to one; a single color corresponds to an unambiguous assignment, several colors to an actor whose co-membership is shared among groups.}\label{fig:rdprof}
\alttext{Stacked bar charts of the normalized membership degrees of each actor to the consensus groups in two departments.}
\end{figure}

\begin{figure}[tbp]
\centerline{\includegraphics[width=0.95\textwidth]{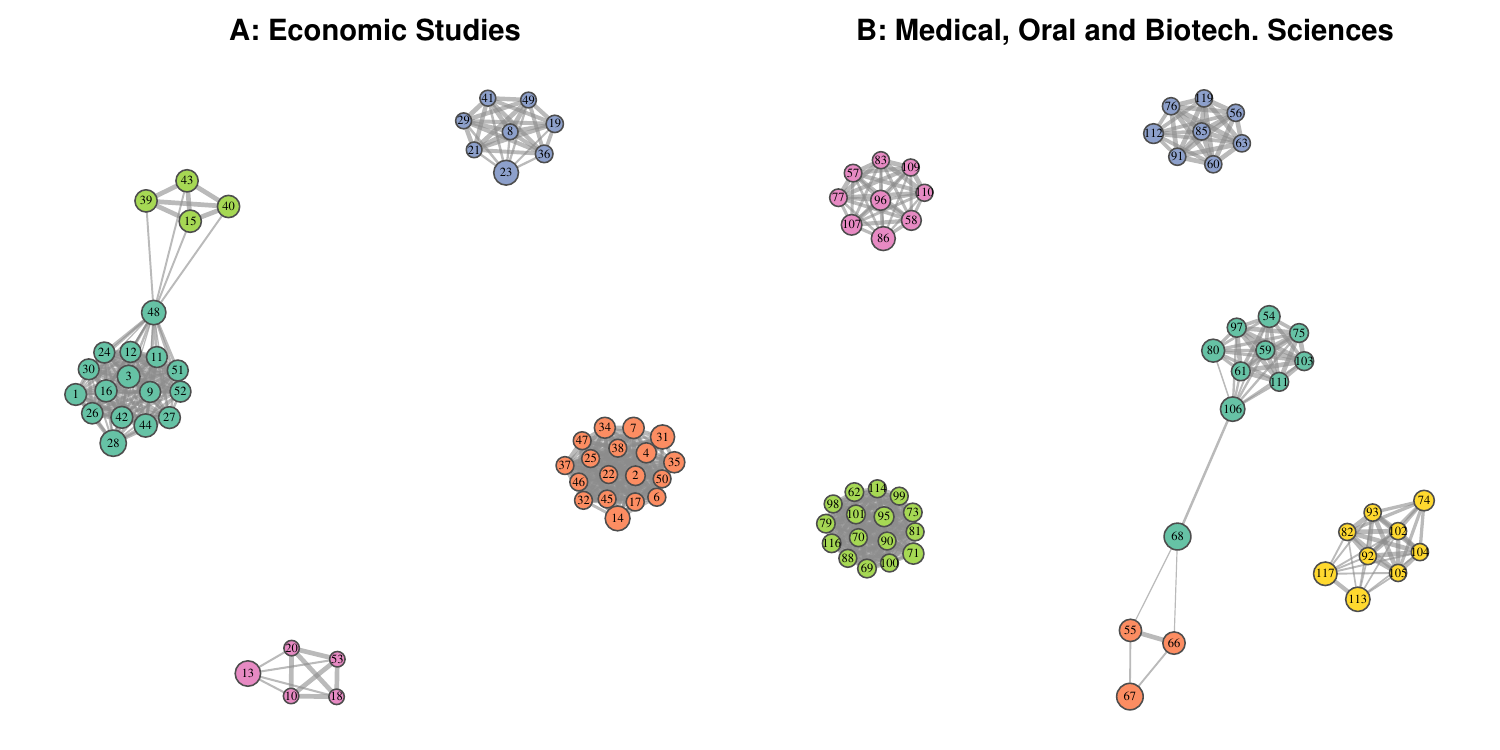}}
\caption{Real networks: consensus networks of departments A (left) and B (right). Edges join pairs of actors with $\mu_{ij}\ge0.5$ and represent co-membership, not the reported collaborations; colors denote FCo-L consensus groups and node size increases with the ambiguity index $\eta_i$, so that large nodes between groups are actors whose assignment is uncertain.}\label{fig:rdnet}
\alttext{Consensus networks of two departments with groups shown by color and node size proportional to ambiguity.}
\end{figure}

\begin{figure}[tbp]
\centerline{\includegraphics[width=0.95\textwidth]{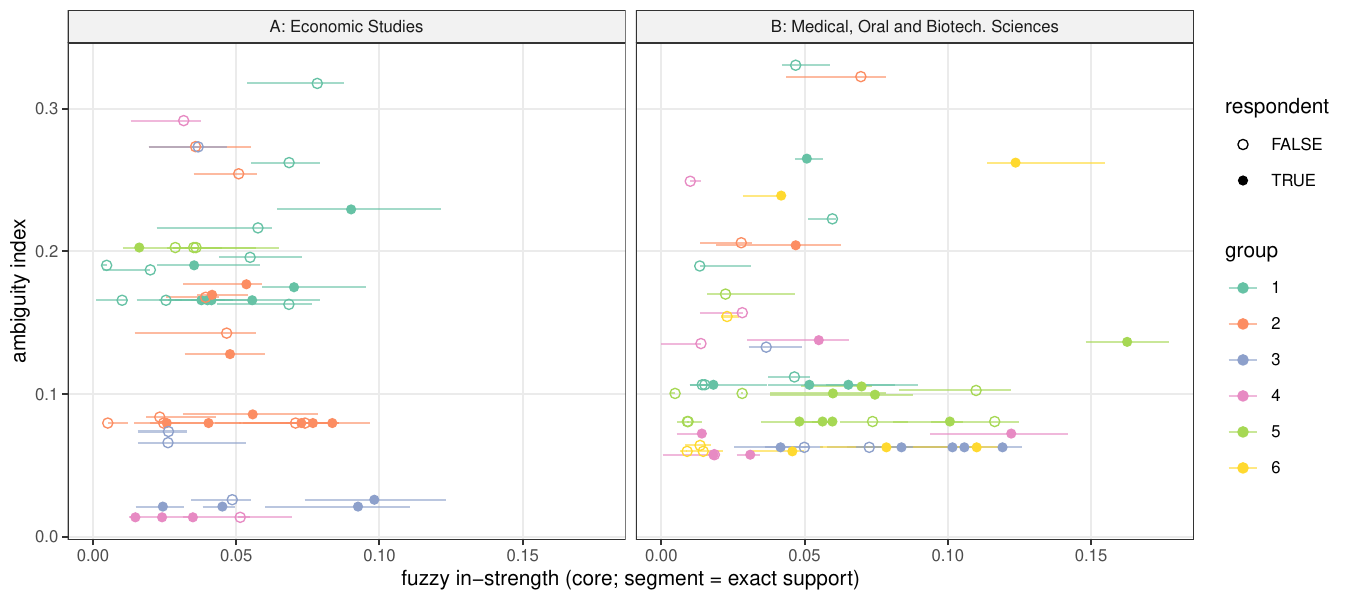}}
\caption{Real networks: ambiguity index $\eta_i$ of FCo-L against the fuzzy in-strength of each actor in departments A (left) and B (right); the point marks the core and the horizontal segment spans the support of the in-strength. Colors denote FCo-L consensus groups; filled points denote respondents, empty points non-respondents. The absence of a trend shows that the ambiguity index does not reproduce centrality.}\label{fig:rdamb}
\alttext{Scatter plots of the ambiguity index against the fuzzy in-strength of each actor in two departments.}
\end{figure}

The relations differ mainly in their degrees. The mean ambiguity is $0.137$ and $0.119$ for FCo-L, $0.154$ and $0.112$ for Perturb-tie and $0.131$ and $0.064$ for Consensus-core; the ranking of the actors by ambiguity is similar under FCo-L and Perturb-tie (Spearman $0.93$ and $0.80$) but not identical. The membership degrees (Fig.~\ref{fig:rdprof}) show that actors with high ambiguity share their membership mostly between two groups. Actor 13 of A, a non-respondent, has mean ambiguity $0.290$ under FCo-L, $0.144$ under Perturb-tie, $0.044$ under Perturb-core and $0$ under Consensus-core: its assignment is stable on the cores and becomes uncertain when the ratings it receives vary within their own asymmetric $\alpha$-cuts. Actor 117 of B behaves similarly ($0.234$, $0.133$, $0.180$ and $0.029$). The benchmarks may also produce ambiguity that FCo-L does not (actor 14 of A: $0.493$ under Perturb-tie, $0.286$ under FCo-L). In the consensus networks (Fig.~\ref{fig:rdnet}) actors with high ambiguity, such as 48 in A and 68 and 106 in B, link otherwise separated groups. The ambiguity index is not a proxy for centrality (Spearman correlation with the core of the fuzzy in-strength $0.04$ and $-0.06$; Fig.~\ref{fig:rdamb}), is moderately correlated with the participation coefficient ($0.27$ and $0.37$) and is lower for respondents than for non-respondents ($0.106$ against $0.165$ in A, $0.105$ against $0.133$ in B). These comparisons are descriptive; no external information on organizational units or research areas is available to validate the groups.

\section{Discussion and Conclusions}\label{sec:disc}

The proposed fuzzy co-membership relation integrates over $\alpha$ the co-assignments of the Louvain algorithm on crisp networks compatible with the $\alpha$-cuts of TFN ties; it needs no preset number of groups, is positive semidefinite and induces a Euclidean distance between actors.

The simulations separate the sources of improvement over a single run on the cores. The largest reduction of the Brier score comes from consensus itself and is shared with the consensus of the cores. Perturbing the ties adds a small improvement of the partition in small networks with weak structure and a further reduction of the Brier score, which are obtained to the same extent by symmetric perturbations of the same size. FCo-L is not a method for obtaining a better partition: the interval-weighted Louvain algorithm\cite{alves2024} is comparable in ARI, better in estimating the number of communities and better at using informative asymmetric supports for the partition.

The fuzzy ties provide a perturbation scale elicited from the respondents rather than tuned by the analyst, a tie-specific distribution of that scale, a nested family of sizes over $\alpha$ and, through the asymmetry of the TFNs, a direction. These features matter beyond a symmetric perturbation of the same size in three situations: with symmetric TFNs the ambiguity index is slightly more closely associated with the structural position of the actors; when the vagueness is concentrated on structurally ambiguous actors, the index identifies them better than both benchmarks; when the asymmetry of the TFNs points towards the latent strength, the co-membership degrees have a lower Brier score, and the difference disappears when the direction is random. The effects are small and arise in designs favorable to FCo-L by construction; whether real ratings are vaguer for structurally ambiguous actors, or asymmetric towards the latent strength, is an empirical question.

In the application FCo-L locates ambiguous actors that neither the consensus of the cores nor a homogeneous perturbation identifies, and the observed ambiguity is lower than in networks with permuted ratings. The partitions, however, depend on the projection of the directed ratings and on the coding of the ties of non-respondents, which is decisive for the placement of non-respondents; the application illustrates the method under these choices and does not validate it. Further limitations concern the design: three balanced blocks, two network sizes, stylized boundary-actor and asymmetric designs, independent realizations across ties, and a lifetime criterion that cannot return a single group. Future work includes the Leiden algorithm,\cite{traag2019} exploration laws reflecting the dependence between ratings of the same respondent, a test of community structure based on $\mu$, and the joint use of the ambiguity index and the fuzzy centrality indices of Porreca et al.\cite{porreca2025} to distinguish actors central within a group from actors bridging groups.

\section*{ORCID}
\noindent Annamaria Porreca - \url{https://orcid.org/0000-0003-3278-1561}

\noindent Fabrizio Maturo - \url{https://orcid.org/0000-0002-2362-4970}

\section*{Declarations}
\noindent\textit{Author contributions.} Both authors contributed to conceptualization, methodology, software, formal analysis, investigation, data curation, validation, visualization, and writing (original draft, review and editing).\\
\textit{Ethical statement.} The questionnaire data were collected by A.\ Porreca during her doctoral research. Participation was voluntary; respondents gave informed consent to the anonymized use of their answers for research and publication.\\
\textit{Funding.} This research did not receive any specific grant from funding agencies in the public, commercial, or not-for-profit sectors.\\
\textit{Competing interests.} The authors declare no competing interests.\\
\textit{Use of generative AI.} The authors used generative AI tools to improve the English language, to assist in revising parts of the text, and to check bibliographic metadata. All content was critically reviewed, verified and edited by the authors, who take full responsibility for the publication.\\
\textit{Data and code availability.} The R script that reproduces all simulations, figures and tables is available from the corresponding author upon request; without the questionnaire data it runs only the simulation study. Results were obtained with R 4.5.2 and igraph 2.3.4 and may differ slightly with other releases, since the Louvain algorithm is randomized. The questionnaire data are available from the corresponding author upon reasonable request.

\clearpage
\appendix
\renewcommand{\thesection}{Appendix \Alph{section}}
\section{Proof of Proposition~\ref{prop:comemb}}\label{app:proofs}
(i) Each indicator in Eq.~(\ref{eq:mu}) is symmetric and equals one on the diagonal, and the weights $w_b$ and $\omega_g$ are non-negative and sum to one. (ii) With crisp ties every box reduces to $\{M\}$, so FCo-L and the control apply $JB$ independent runs of $\Cl$ to the same network with the same weights; with a deterministic algorithm all indicators coincide. (iii) Relabeling the actors by a permutation matrix $\Pi$ maps $\W_\alpha$ onto $\{\Pi W\Pi^\top:W\in\W_\alpha\}$; the deterministic realizations are permuted exactly, and the uniform ones have the same joint distribution after permutation since the $U^{(b)}_{ij}$ are i.i.d. By equivariance in distribution of $\Cl$ and independence of the runs, each co-membership matrix is permuted in distribution, hence so are $\pi(\alpha_g)$ and $\mu$, which are fixed linear combinations of independent co-membership matrices; taking expectations, $\mathrm{E}(\mu)$ is permuted exactly. A randomized algorithm such as Louvain is equivariant only in distribution: for a fixed seed, relabeling the actors changes the visiting order. (iv) The co-membership matrix of a partition with indicator matrix $Z\in\{0,1\}^{n\times K}$ is $ZZ^\top\succeq0$ with unit diagonal, and $\mu$ is a convex combination of such matrices; hence $\mu\succeq0$ and $\mu_{ii}=1$. A positive semidefinite matrix is a Gram matrix, $\mu_{ij}=\langle x_i,x_j\rangle$, and $\|x_i-x_j\|^2=2(1-\mu_{ij})$; $d_{ij}=0$ for $i\ne j$ only if $\mu_{ij}=1$, in which case $d$ is a pseudometric.

\clearpage
\setcounter{table}{0}
\renewcommand{\thetable}{B\arabic{table}}
\begin{landscape}
\section{Detailed Simulation Comparisons}\label{app:tables}

\begin{table}[H]
\centering
\tbl{Main simulation design, FCo-L minus Consensus-core, paired over $100$ replicates per scenario: ARI, mean (standard error), $z$ ratio and number of replicates in which FCo-L is better, equal or worse (win/tie/loss); Brier score; AUC of the ambiguity index for each relation's own misclassified actors, with the number of replicates in which both are defined in brackets; structural Spearman correlation. Positive differences favor FCo-L except for the Brier score.\label{tab:core}}
{\adjustbox{max width=\linewidth,max totalheight=0.16\textheight}{\input{tables/Table_B1_Paired_FCoL_vs_Core.tex}}}
\end{table}

\begin{table}[H]
\centering
\tbl{Main simulation design, FCo-L minus Perturb-core (common perturbation size with matched mean absolute deviation); same quantities as Table~\ref{tab:core}.\label{tab:pert}}
{\adjustbox{max width=\linewidth,max totalheight=0.16\textheight}{\input{tables/Table_B2_Paired_FCoL_vs_Perturb.tex}}}
\end{table}

\begin{table}[H]
\centering
\tbl{Main simulation design, FCo-L minus Perturb-tie (tie-specific symmetric perturbation with matched mean absolute deviation); same quantities as Table~\ref{tab:core}.\label{tab:tie}}
{\adjustbox{max width=\linewidth,max totalheight=0.16\textheight}{\input{tables/Table_B3_Paired_FCoL_vs_PerturbTie.tex}}}
\end{table}

\end{landscape}

\begin{table}[p]
\centering
\tbl{Asymmetric-TFN design, $100$ replicates per scenario: ARI, mean (standard error); proportion of replicates with three communities; Brier score; structural Spearman correlation; AUC of the ambiguity index for misclassified actors, with the number of replicates in which it is defined in brackets; share of ties with $|(r-m)-(m-l)|>(r-l)/4$. \emph{Informative}: the longer side of each TFN points towards the latent strength; \emph{random\_direction}: left and right spreads swapped with probability $1/2$; \emph{common}: methods that use only the cores.\label{tab:asym}}
{\adjustbox{max width=\textwidth,max totalheight=0.85\textheight}{\input{tables/Table_B4_Asymmetric_Summary.tex}}}
\end{table}

\begin{table}[p]
\centering
\tbl{Asymmetric-TFN design: paired differences on the same networks, mean (standard error) and $z$ ratio, of FCo-L against each benchmark within each arm, and of FCo-L and of Perturb-tie between the informative and the random-direction arm. For the Brier score negative values favor FCo-L (within arms) or the informative arm (between arms).\label{tab:asympaired}}
{\adjustbox{max width=\textwidth,max totalheight=0.85\textheight}{\input{tables/Table_B5_Asymmetric_Paired.tex}}}
\end{table}

\end{document}

%% file: tables/Table_01_Simulation_Recovery_CoMembership.tex
\begin{tabular}{lrrrrrrrr}
\toprule
method & ARI & NMI & $\hat K=3$ & mean $\hat K$ & Brier & AUC pairs & AUC $\eta$, misclassified & structural Spearman \\
\midrule
\multicolumn{9}{l}{\textit{$n=30$, $p_{\mathrm{out}}=0.05$, informative spreads}} \\
FCo-L & 0.877 (0.014) & 0.895 (0.011) & 0.75 & 3.35 & 0.042 (0.004) & 0.976 & 0.856 [55] & 0.313 (0.019) \\
Consensus-core & 0.869 (0.014) & 0.890 (0.011) & 0.67 & 3.43 & 0.046 (0.005) & 0.963 & 0.789 [54] & 0.237 (0.019) \\
Perturb-core & 0.871 (0.014) & 0.890 (0.011) & 0.70 & 3.41 & 0.043 (0.005) & 0.973 & 0.842 [55] & 0.286 (0.018) \\
Perturb-tie & 0.879 (0.013) & 0.897 (0.010) & 0.71 & 3.40 & 0.042 (0.004) & 0.976 & 0.835 [55] & 0.302 (0.019) \\
Louvain-core & 0.856 (0.015) & 0.877 (0.012) & 0.68 & 3.40 & 0.059 (0.006) & 0.922 & -- & -- \\
Louvain-CoG & 0.859 (0.015) & 0.881 (0.011) & 0.73 & 3.35 & 0.058 (0.006) & 0.924 & -- & -- \\
IW-Louvain & 0.863 (0.015) & 0.884 (0.012) & 0.84 & 3.18 & 0.059 (0.006) & 0.931 & -- & -- \\
\addlinespace
\multicolumn{9}{l}{\textit{$n=60$, $p_{\mathrm{out}}=0.05$, informative spreads}} \\
FCo-L & 0.994 (0.002) & 0.993 (0.002) & 0.99 & 3.01 & 0.002 (0.001) & 0.999 & 0.942 [10] & 0.123 (0.013) \\
Consensus-core & 0.994 (0.002) & 0.993 (0.002) & 0.98 & 3.02 & 0.002 (0.001) & 0.999 & 0.936 [9] & 0.105 (0.016) \\
Perturb-core & 0.994 (0.002) & 0.993 (0.002) & 0.98 & 3.02 & 0.002 (0.001) & 0.999 & 0.942 [10] & 0.121 (0.016) \\
Perturb-tie & 0.994 (0.002) & 0.993 (0.002) & 0.97 & 3.03 & 0.002 (0.001) & 0.999 & 0.936 [9] & 0.115 (0.015) \\
Louvain-core & 0.992 (0.002) & 0.991 (0.003) & 0.99 & 3.01 & 0.003 (0.001) & 0.996 & -- & -- \\
Louvain-CoG & 0.993 (0.002) & 0.991 (0.002) & 0.99 & 3.01 & 0.003 (0.001) & 0.996 & -- & -- \\
IW-Louvain & 0.993 (0.002) & 0.992 (0.002) & 1.00 & 3.00 & 0.003 (0.001) & 0.997 & -- & -- \\
\addlinespace
\multicolumn{9}{l}{\textit{$n=30$, $p_{\mathrm{out}}=0.10$, informative spreads}} \\
FCo-L & 0.698 (0.019) & 0.744 (0.016) & 0.46 & 3.67 & 0.094 (0.005) & 0.928 & 0.857 [83] & 0.438 (0.016) \\
Consensus-core & 0.671 (0.019) & 0.724 (0.015) & 0.36 & 3.87 & 0.102 (0.006) & 0.916 & 0.824 [85] & 0.366 (0.019) \\
Perturb-core & 0.693 (0.020) & 0.741 (0.016) & 0.47 & 3.69 & 0.094 (0.006) & 0.925 & 0.844 [83] & 0.399 (0.017) \\
Perturb-tie & 0.694 (0.019) & 0.741 (0.016) & 0.44 & 3.74 & 0.093 (0.005) & 0.929 & 0.843 [83] & 0.411 (0.017) \\
Louvain-core & 0.671 (0.019) & 0.726 (0.016) & 0.37 & 3.79 & 0.133 (0.008) & 0.824 & -- & -- \\
Louvain-CoG & 0.673 (0.020) & 0.725 (0.016) & 0.41 & 3.67 & 0.133 (0.008) & 0.827 & -- & -- \\
IW-Louvain & 0.667 (0.024) & 0.714 (0.021) & 0.81 & 2.86 & 0.152 (0.011) & 0.843 & -- & -- \\
\addlinespace
\multicolumn{9}{l}{\textit{$n=60$, $p_{\mathrm{out}}=0.10$, informative spreads}} \\
FCo-L & 0.972 (0.004) & 0.968 (0.005) & 0.97 & 3.03 & 0.010 (0.001) & 0.998 & 0.954 [38] & 0.339 (0.013) \\
Consensus-core & 0.970 (0.005) & 0.966 (0.005) & 0.99 & 3.01 & 0.010 (0.001) & 0.998 & 0.973 [41] & 0.246 (0.015) \\
Perturb-core & 0.971 (0.004) & 0.967 (0.004) & 0.95 & 3.06 & 0.010 (0.001) & 0.998 & 0.938 [39] & 0.284 (0.014) \\
Perturb-tie & 0.971 (0.004) & 0.967 (0.005) & 0.98 & 3.03 & 0.010 (0.001) & 0.998 & 0.948 [40] & 0.310 (0.014) \\
Louvain-core & 0.956 (0.005) & 0.950 (0.006) & 0.98 & 3.02 & 0.019 (0.002) & 0.978 & -- & -- \\
Louvain-CoG & 0.951 (0.006) & 0.945 (0.006) & 0.97 & 3.03 & 0.021 (0.003) & 0.976 & -- & -- \\
IW-Louvain & 0.967 (0.006) & 0.964 (0.005) & 1.00 & 3.00 & 0.015 (0.003) & 0.984 & -- & -- \\
\addlinespace
\multicolumn{9}{l}{\textit{$n=30$, $p_{\mathrm{out}}=0.05$, uninformative spreads}} \\
FCo-L & 0.883 (0.012) & 0.898 (0.010) & 0.74 & 3.32 & 0.040 (0.004) & 0.976 & 0.845 [57] & 0.339 (0.016) \\
Consensus-core & 0.877 (0.013) & 0.893 (0.010) & 0.72 & 3.33 & 0.043 (0.004) & 0.967 & 0.701 [59] & 0.210 (0.024) \\
Perturb-core & 0.880 (0.012) & 0.896 (0.010) & 0.74 & 3.35 & 0.040 (0.004) & 0.975 & 0.824 [57] & 0.279 (0.019) \\
Perturb-tie & 0.876 (0.013) & 0.893 (0.010) & 0.72 & 3.35 & 0.040 (0.004) & 0.975 & 0.823 [58] & 0.304 (0.017) \\
Louvain-core & 0.873 (0.013) & 0.888 (0.011) & 0.73 & 3.28 & 0.053 (0.005) & 0.932 & -- & -- \\
Louvain-CoG & 0.879 (0.012) & 0.893 (0.010) & 0.69 & 3.32 & 0.050 (0.005) & 0.935 & -- & -- \\
IW-Louvain & 0.853 (0.015) & 0.870 (0.013) & 0.82 & 3.17 & 0.063 (0.007) & 0.926 & -- & -- \\
\addlinespace
\multicolumn{9}{l}{\textit{$n=60$, $p_{\mathrm{out}}=0.05$, uninformative spreads}} \\
FCo-L & 0.997 (0.001) & 0.997 (0.001) & 0.98 & 3.02 & 0.001 (0.000) & 1.000 & 0.983 [4] & 0.087 (0.014) \\
Consensus-core & 0.997 (0.001) & 0.997 (0.001) & 0.97 & 3.03 & 0.001 (0.000) & 0.999 & 0.739 [4] & 0.069 (0.015) \\
Perturb-core & 0.997 (0.001) & 0.997 (0.001) & 0.98 & 3.02 & 0.001 (0.000) & 1.000 & 0.987 [4] & 0.084 (0.012) \\
Perturb-tie & 0.997 (0.001) & 0.997 (0.001) & 0.98 & 3.02 & 0.001 (0.000) & 1.000 & 0.987 [4] & 0.090 (0.015) \\
Louvain-core & 0.994 (0.002) & 0.993 (0.002) & 1.00 & 3.00 & 0.002 (0.001) & 0.997 & -- & -- \\
Louvain-CoG & 0.995 (0.002) & 0.994 (0.002) & 1.00 & 3.00 & 0.002 (0.001) & 0.997 & -- & -- \\
IW-Louvain & 0.994 (0.002) & 0.993 (0.002) & 1.00 & 3.00 & 0.002 (0.001) & 0.997 & -- & -- \\
\addlinespace
\multicolumn{9}{l}{\textit{$n=30$, $p_{\mathrm{out}}=0.10$, uninformative spreads}} \\
FCo-L & 0.680 (0.016) & 0.733 (0.013) & 0.38 & 3.85 & 0.099 (0.004) & 0.923 & 0.836 [92] & 0.407 (0.017) \\
Consensus-core & 0.667 (0.016) & 0.726 (0.013) & 0.33 & 4.00 & 0.108 (0.005) & 0.908 & 0.788 [90] & 0.302 (0.020) \\
Perturb-core & 0.690 (0.016) & 0.744 (0.013) & 0.38 & 3.94 & 0.099 (0.005) & 0.923 & 0.818 [90] & 0.375 (0.017) \\
Perturb-tie & 0.692 (0.016) & 0.744 (0.013) & 0.44 & 3.80 & 0.097 (0.004) & 0.925 & 0.842 [92] & 0.403 (0.015) \\
Louvain-core & 0.648 (0.017) & 0.711 (0.013) & 0.36 & 3.89 & 0.142 (0.007) & 0.811 & -- & -- \\
Louvain-CoG & 0.643 (0.018) & 0.702 (0.014) & 0.34 & 3.81 & 0.145 (0.007) & 0.810 & -- & -- \\
IW-Louvain & 0.682 (0.022) & 0.724 (0.019) & 0.83 & 2.93 & 0.144 (0.011) & 0.847 & -- & -- \\
\addlinespace
\multicolumn{9}{l}{\textit{$n=60$, $p_{\mathrm{out}}=0.10$, uninformative spreads}} \\
FCo-L & 0.968 (0.005) & 0.964 (0.006) & 0.92 & 3.08 & 0.011 (0.001) & 0.997 & 0.944 [37] & 0.337 (0.014) \\
Consensus-core & 0.969 (0.005) & 0.965 (0.005) & 0.92 & 3.08 & 0.011 (0.002) & 0.996 & 0.893 [35] & 0.248 (0.014) \\
Perturb-core & 0.970 (0.005) & 0.966 (0.005) & 0.93 & 3.07 & 0.011 (0.001) & 0.997 & 0.914 [34] & 0.282 (0.013) \\
Perturb-tie & 0.969 (0.005) & 0.964 (0.005) & 0.92 & 3.08 & 0.011 (0.001) & 0.997 & 0.932 [36] & 0.305 (0.014) \\
Louvain-core & 0.951 (0.007) & 0.945 (0.007) & 0.95 & 3.05 & 0.021 (0.003) & 0.975 & -- & -- \\
Louvain-CoG & 0.945 (0.007) & 0.939 (0.007) & 0.92 & 3.08 & 0.024 (0.003) & 0.971 & -- & -- \\
IW-Louvain & 0.967 (0.005) & 0.963 (0.005) & 1.00 & 3.00 & 0.015 (0.002) & 0.984 & -- & -- \\
\bottomrule
\end{tabular}

%% file: tables/Table_02_Paired_Differences.tex
\begin{tabular}{lllllll}
\toprule
scenario & vs Consensus-core & vs Perturb-core & vs Perturb-tie & vs Louvain-core & vs Louvain-CoG & vs IW-Louvain \\
\midrule
n=30, p\_out=0.05, informative & 0.008 (0.005) & 0.006 (0.005) & -0.001 (0.004) & 0.021 (0.007) & 0.018 (0.007) & 0.014 (0.011) \\
n=60, p\_out=0.05, informative & -0.000 (0.001) & 0.000 (0.000) & -0.000 (0.000) & 0.002 (0.002) & 0.001 (0.001) & 0.001 (0.002) \\
n=30, p\_out=0.10, informative & 0.027 (0.007) & 0.005 (0.007) & 0.003 (0.006) & 0.026 (0.013) & 0.025 (0.011) & 0.030 (0.018) \\
n=60, p\_out=0.10, informative & 0.002 (0.001) & 0.001 (0.002) & 0.001 (0.001) & 0.016 (0.004) & 0.020 (0.004) & 0.005 (0.006) \\
n=30, p\_out=0.05, uninformative & 0.006 (0.005) & 0.003 (0.003) & 0.007 (0.005) & 0.011 (0.007) & 0.004 (0.006) & 0.030 (0.012) \\
n=60, p\_out=0.05, uninformative & 0.000 (0.000) & 0.000 (0.000) & 0.000 (0.000) & 0.003 (0.001) & 0.003 (0.001) & 0.003 (0.001) \\
n=30, p\_out=0.10, uninformative & 0.013 (0.007) & -0.010 (0.008) & -0.012 (0.005) & 0.032 (0.011) & 0.037 (0.010) & -0.002 (0.019) \\
n=60, p\_out=0.10, uninformative & -0.001 (0.001) & -0.002 (0.001) & -0.000 (0.001) & 0.018 (0.005) & 0.024 (0.006) & 0.002 (0.006) \\
\bottomrule
\end{tabular}

%% file: tables/Table_03_Boundary_Summary.tex
\begin{tabular}{llllllll}
\toprule
scenario & arm & method & ARI & AUC boundary & AUC misclassified & ambiguity boundary/other & misclassified boundary/other \\
\midrule
boundary, n=30 & aligned & FCo-L & 0.671 (0.014) & 0.807 (0.014) & 0.823 [93] & 0.215/0.101 & 0.51/0.05 \\
boundary, n=30 & aligned & Perturb-core & 0.669 (0.013) & 0.725 (0.018) & 0.773 [94] & 0.179/0.091 & 0.50/0.05 \\
boundary, n=30 & aligned & Perturb-tie & 0.672 (0.013) & 0.783 (0.017) & 0.810 [93] & 0.215/0.101 & 0.51/0.05 \\
boundary, n=30 & misaligned & FCo-L & 0.665 (0.013) & 0.749 (0.015) & 0.794 [94] & 0.186/0.098 & 0.51/0.05 \\
boundary, n=30 & misaligned & Perturb-core & 0.672 (0.013) & 0.735 (0.018) & 0.790 [94] & 0.179/0.092 & 0.52/0.05 \\
boundary, n=30 & misaligned & Perturb-tie & 0.676 (0.013) & 0.737 (0.017) & 0.785 [94] & 0.188/0.099 & 0.52/0.05 \\
boundary, n=30 & common & Consensus-core & 0.667 (0.013) & 0.670 (0.018) & 0.711 [94] & 0.134/0.065 & 0.51/0.05 \\
boundary, n=30 & common & Louvain-core & 0.654 (0.015) & -- & -- & -- & -- \\
boundary, n=60 & aligned & FCo-L & 0.857 (0.007) & 0.874 (0.009) & 0.865 [94] & 0.191/0.030 & 0.46/0.00 \\
boundary, n=60 & aligned & Perturb-core & 0.856 (0.007) & 0.811 (0.011) & 0.808 [94] & 0.166/0.026 & 0.46/0.00 \\
boundary, n=60 & aligned & Perturb-tie & 0.854 (0.007) & 0.855 (0.010) & 0.863 [94] & 0.196/0.030 & 0.47/0.00 \\
boundary, n=60 & misaligned & FCo-L & 0.855 (0.007) & 0.826 (0.011) & 0.837 [94] & 0.171/0.028 & 0.46/0.00 \\
boundary, n=60 & misaligned & Perturb-core & 0.855 (0.007) & 0.794 (0.012) & 0.800 [94] & 0.165/0.026 & 0.46/0.00 \\
boundary, n=60 & misaligned & Perturb-tie & 0.854 (0.007) & 0.813 (0.011) & 0.824 [94] & 0.169/0.027 & 0.47/0.00 \\
boundary, n=60 & common & Consensus-core & 0.855 (0.007) & 0.774 (0.012) & 0.771 [95] & 0.137/0.022 & 0.46/0.00 \\
boundary, n=60 & common & Louvain-core & 0.847 (0.007) & -- & -- & -- & -- \\
\bottomrule
\end{tabular}

%% file: tables/Table_04_Boundary_Paired.tex
\begin{tabular}{lllllll}
\toprule
scenario & arm & comparison & ARI & ARI z & AUC boundary & AUC z \\
\midrule
boundary, n=30 & aligned & FCo-L minus Consensus-core & 0.005 (0.007) & 0.67 & 0.137 (0.014) & 9.75 \\
boundary, n=30 & aligned & FCo-L minus Perturb-core & 0.003 (0.004) & 0.60 & 0.082 (0.010) & 7.91 \\
boundary, n=30 & aligned & FCo-L minus Perturb-tie & -0.000 (0.004) & -0.11 & 0.025 (0.007) & 3.48 \\
boundary, n=30 & misaligned & FCo-L minus Consensus-core & -0.001 (0.006) & -0.26 & 0.079 (0.013) & 6.05 \\
boundary, n=30 & misaligned & FCo-L minus Perturb-core & -0.006 (0.005) & -1.28 & 0.014 (0.008) & 1.74 \\
boundary, n=30 & misaligned & FCo-L minus Perturb-tie & -0.010 (0.004) & -2.55 & 0.012 (0.007) & 1.80 \\
boundary, n=60 & aligned & FCo-L minus Consensus-core & 0.002 (0.002) & 0.81 & 0.101 (0.010) & 9.86 \\
boundary, n=60 & aligned & FCo-L minus Perturb-core & 0.001 (0.002) & 0.73 & 0.063 (0.009) & 7.24 \\
boundary, n=60 & aligned & FCo-L minus Perturb-tie & 0.003 (0.002) & 1.58 & 0.019 (0.006) & 3.01 \\
boundary, n=60 & misaligned & FCo-L minus Consensus-core & -0.000 (0.002) & -0.15 & 0.052 (0.009) & 5.61 \\
boundary, n=60 & misaligned & FCo-L minus Perturb-core & -0.000 (0.002) & -0.22 & 0.032 (0.009) & 3.74 \\
boundary, n=60 & misaligned & FCo-L minus Perturb-tie & 0.001 (0.002) & 0.50 & 0.012 (0.006) & 2.00 \\
\bottomrule
\end{tabular}

%% file: tables/Table_05_Sensitivity.tex
\begin{tabular}{lllllll}
\toprule
scenario & variant & ARI & K\_correct & K\_mean & Brier & AUC\_ambiguity \\
\midrule
informative & default (mixed, B=25, step 0.2, K<=8) & 0.698 (0.019) & 0.46 & 3.67 & 0.0938 (0.0054) & 0.857 [83] \\
informative & uniform realizations only & 0.695 (0.020) & 0.49 & 3.68 & 0.0939 (0.0055) & 0.857 [83] \\
informative & B=50 & 0.696 (0.020) & 0.48 & 3.65 & 0.0934 (0.0055) & 0.857 [83] \\
informative & alpha step 0.1 & 0.694 (0.020) & 0.46 & 3.66 & 0.0935 (0.0055) & 0.861 [83] \\
informative & K<=15 & 0.698 (0.019) & 0.46 & 3.67 & 0.0938 (0.0054) & 0.857 [83] \\
uninformative & default (mixed, B=25, step 0.2, K<=8) & 0.680 (0.016) & 0.38 & 3.85 & 0.0992 (0.0045) & 0.836 [92] \\
uninformative & uniform realizations only & 0.682 (0.016) & 0.40 & 3.91 & 0.0993 (0.0045) & 0.826 [91] \\
uninformative & B=50 & 0.685 (0.016) & 0.37 & 3.92 & 0.0988 (0.0045) & 0.831 [92] \\
uninformative & alpha step 0.1 & 0.679 (0.016) & 0.38 & 3.95 & 0.0991 (0.0045) & 0.825 [90] \\
uninformative & K<=15 & 0.680 (0.016) & 0.38 & 3.85 & 0.0992 (0.0045) & 0.836 [92] \\
\bottomrule
\end{tabular}

%% file: tables/Table_06_Null_Scenario.tex
\begin{tabular}{lllllll}
\toprule
scenario & method & share K=2 & K\_mean & mean co-membership & mean ambiguity (null) & mean ambiguity (block model) \\
\midrule
null, n=30 & FCo-L & 0.03 & 4.53 & 0.222 & 0.258 & 0.090 \\
null, n=30 & Consensus-core & 0.00 & 4.56 & 0.222 & 0.211 & 0.062 \\
null, n=30 & Perturb-core & 0.02 & 4.47 & 0.222 & 0.245 & 0.082 \\
null, n=30 & Perturb-tie & 0.03 & 4.39 & 0.222 & 0.257 & 0.089 \\
null, n=30 & Louvain-core & 0.00 & 4.47 & 0.224 & -- & -- \\
null, n=30 & Louvain-CoG & 0.00 & 4.43 & 0.223 & -- & -- \\
null, n=30 & IW-Louvain & 0.35 & 2.68 & 0.400 & -- & -- \\
null, n=60 & FCo-L & 0.02 & 5.44 & 0.187 & 0.307 & 0.018 \\
null, n=60 & Consensus-core & 0.01 & 5.44 & 0.186 & 0.293 & 0.014 \\
null, n=60 & Perturb-core & 0.01 & 5.44 & 0.186 & 0.300 & 0.016 \\
null, n=60 & Perturb-tie & 0.03 & 5.22 & 0.186 & 0.305 & 0.018 \\
null, n=60 & Louvain-core & 0.00 & 5.55 & 0.187 & -- & -- \\
null, n=60 & Louvain-CoG & 0.00 & 5.48 & 0.189 & -- & -- \\
null, n=60 & IW-Louvain & 0.10 & 2.96 & 0.379 & -- & -- \\
\bottomrule
\end{tabular}

%% file: tables/Table_RealData_01_Summary.tex
\begin{tabular}{lll}
\toprule
quantity & A: Economic Studies & B: Medical, Oral and Biotech. Sciences \\
\midrule
actors & 51 & 55 \\
respondents & 24 & 28 \\
directed ties & 219 & 254 \\
groups, FCo-L & 5 & 6 \\
groups, Consensus-core & 4 & 6 \\
groups, Louvain-core & 4 & 6 \\
groups, IW-Louvain & 5 & 4 \\
median groups over realizations & 5 & 5 \\
FCo-L group sizes & 16/18/8/5/4 & 10/3/8/9/16/9 \\
mean ambiguity, FCo-L & 0.137 & 0.119 \\
mean ambiguity, Consensus-core & 0.131 & 0.064 \\
mean ambiguity, Perturb-core & 0.146 & 0.108 \\
mean ambiguity, Perturb-tie & 0.154 & 0.112 \\
delta, Perturb-core & 0.139 & 0.113 \\
share of ties with core 1 & 0.032 & 0.059 \\
share of ties with final position at button 1 & 0.005 & 0.004 \\
mean support width r - l & 0.313 & 0.254 \\
share of ties with |(r-m)-(m-l)| > (r-l)/4 & 0.936 & 0.839 \\
mean (r-m)-(m-l) & -0.016 & -0.059 \\
\bottomrule
\end{tabular}

%% file: tables/Table_RealData_02_Agreement.tex
\begin{tabular}{lllllll}
\toprule
department & method & groups & ARI & NMI & ARI over repetitions & groups over repetitions \\
\midrule
A & Consensus-core & 4 & 0.827 & 0.876 & 0.839 [0.827, 0.948] & 4--5 \\
A & Perturb-core & 5 & 0.948 & 0.953 & 0.936 [0.827, 0.948] & 4--5 \\
A & Louvain-core & 4 & 0.616 & 0.778 & 0.805 [0.589, 0.948] & 4--5 \\
A & Louvain-CoG & 5 & 0.931 & 0.948 & 0.869 [0.525, 1.000] & 4--5 \\
A & IW-Louvain & 5 & 0.768 & 0.808 & 0.569 [0.317, 0.768] & 3--5 \\
B & Consensus-core & 6 & 1.000 & 1.000 & 1.000 [1.000, 1.000] & 6 \\
B & Perturb-core & 6 & 0.887 & 0.912 & 0.882 [0.836, 0.887] & 5--7 \\
B & Louvain-core & 6 & 1.000 & 1.000 & 0.863 [0.564, 1.000] & 4--7 \\
B & Louvain-CoG & 5 & 0.835 & 0.884 & 0.798 [0.625, 0.972] & 4--7 \\
B & IW-Louvain & 4 & 0.623 & 0.739 & 0.525 [0.374, 0.628] & 3--5 \\
\bottomrule
\end{tabular}

%% file: tables/Table_RealData_03_Permutation_Reference.tex
\begin{tabular}{llllll}
\toprule
department & permutations & mean ambiguity, observed & mean ambiguity, permuted & groups, observed / permuted & mean co-membership, observed / permuted \\
\midrule
A & 20 & 0.137 & 0.247 [0.213, 0.297] & 5 / 6.00 & 0.250 / 0.192 \\
B & 20 & 0.119 & 0.250 [0.212, 0.306] & 6 / 5.80 & 0.200 / 0.181 \\
\bottomrule
\end{tabular}

%% file: tables/Table_RealData_04_Stability.tex
\begin{tabular}{lll}
\toprule
quantity & A: Economic Studies & B: Medical, Oral and Biotech. Sciences \\
\midrule
independent repetitions & 20 & 20 \\
SD of $\mu_{ij}$ over repetitions, FCo-L: mean & 0.012 & 0.011 \\
SD of $\mu_{ij}$, FCo-L: max & 0.048 & 0.049 \\
SD of $\mu^{\mathrm{core}}_{ij}$: mean & 0.011 & 0.008 \\
SD of $\mu^{\mathrm{core}}_{ij}$: max & 0.056 & 0.044 \\
SD of $\eta_i$, FCo-L: max over actors & 0.033 & 0.031 \\
ARI between repetitions, FCo-L: mean & 0.940 & 0.961 \\
ARI between repetitions, FCo-L: min & 0.874 & 0.865 \\
ARI between repetitions, Consensus-core: mean & 0.977 & 1 \\
ARI between repetitions, Consensus-core: min & 0.877 & 1 \\
share of repetitions with the modal $\hat K$, FCo-L & 0.650 & 0.750 \\
actors with $\eta_i$ higher under FCo-L ($z>2$) & 27 & 55 \\
actors with $\eta_i$ lower under FCo-L ($z<-2$) & 11 & 0 \\
groups, FCo-L with $\Phi=$ mean & 5 & 5 \\
ARI, FCo-L with $\Phi=$ mean vs $\Phi=\max$ & 0.640 & 0.629 \\
Spearman, ambiguity vs participation coefficient & 0.274 & 0.365 \\
Spearman, ambiguity FCo-L vs Perturb-core (repetition means) & 0.870 & 0.714 \\
Spearman, ambiguity FCo-L vs Perturb-tie (repetition means) & 0.927 & 0.795 \\
\bottomrule
\end{tabular}

%% file: tables/Table_RealData_05_Unknown_Ties.tex
\begin{tabular}{lllllll}
\toprule
department & convention & pairs unknown & groups & ARI vs main & mean ambiguity & respondents / non-respondents \\
\midrule
A & null ties (main analysis) & 0.00 & 5 & 1.000 & 0.137 & 0.106 / 0.165 \\
A & nonrespondent pairs & 0.28 & 2 & -0.025 & 0.143 & 0.189 / 0.102 \\
A & all ties sent by non-respondents & 0.78 & 2 & 0.317 & 0.468 & 0.427 / 0.504 \\
A & respondents only (24 actors) & -- & 4 & 0.933 & 0.107 & 0.107 / -- \\
B & null ties (main analysis) & 0.00 & 6 & 1.000 & 0.119 & 0.105 / 0.133 \\
B & nonrespondent pairs & 0.24 & 3 & 0.057 & 0.136 & 0.206 / 0.064 \\
B & all ties sent by non-respondents & 0.75 & 4 & 0.301 & 0.418 & 0.362 / 0.475 \\
B & respondents only (28 actors) & -- & 4 & 0.712 & 0.009 & 0.009 / -- \\
\bottomrule
\end{tabular}

%% file: tables/Table_B1_Paired_FCoL_vs_Core.tex
\begin{tabular}{llllllllll}
\toprule
scenario & ARI & ARI z & win/tie/loss & Brier & Brier z & ambiguity AUC & AUC z & structural Spearman & Spearman z \\
\midrule
n=30, p\_out=0.05, informative & 0.008 (0.005) & 1.74 & 12/84/4 & -0.0046 (0.0011) & -4.24 & 0.068 (0.031) [51] & 2.22 & 0.086 (0.017) & 5.10 \\
n=60, p\_out=0.05, informative & -0.000 (0.001) & -0.45 & 1/98/1 & 0.0000 (0.0001) & 0.11 & -0.000 (0.000) [9] & -1.00 & 0.029 (0.017) & 1.65 \\
n=30, p\_out=0.10, informative & 0.027 (0.007) & 3.84 & 26/68/6 & -0.0080 (0.0014) & -5.74 & 0.037 (0.017) [83] & 2.14 & 0.079 (0.015) & 5.36 \\
n=60, p\_out=0.10, informative & 0.002 (0.001) & 1.48 & 6/93/1 & -0.0000 (0.0003) & -0.11 & -0.018 (0.014) [37] & -1.26 & 0.093 (0.013) & 7.11 \\
n=30, p\_out=0.05, uninformative & 0.006 (0.005) & 1.35 & 8/87/5 & -0.0027 (0.0008) & -3.26 & 0.154 (0.035) [57] & 4.35 & 0.141 (0.020) & 7.06 \\
n=60, p\_out=0.05, uninformative & 0.000 (0.000) & 1.00 & 1/99/0 & -0.0001 (0.0001) & -1.00 & 0.244 (0.200) [4] & 1.22 & 0.031 (0.017) & 1.80 \\
n=30, p\_out=0.10, uninformative & 0.013 (0.007) & 1.81 & 24/60/16 & -0.0092 (0.0012) & -7.76 & 0.047 (0.015) [89] & 3.03 & 0.105 (0.016) & 6.63 \\
n=60, p\_out=0.10, uninformative & -0.001 (0.001) & -0.80 & 3/93/4 & 0.0001 (0.0002) & 0.63 & 0.049 (0.022) [34] & 2.25 & 0.092 (0.010) & 9.07 \\
\bottomrule
\end{tabular}

%% file: tables/Table_B2_Paired_FCoL_vs_Perturb.tex
\begin{tabular}{llllllllll}
\toprule
scenario & ARI & ARI z & win/tie/loss & Brier & Brier z & ambiguity AUC & AUC z & structural Spearman & Spearman z \\
\midrule
n=30, p\_out=0.05, informative & 0.006 (0.005) & 1.27 & 12/85/3 & -0.0011 (0.0006) & -1.86 & 0.015 (0.015) [53] & 0.99 & 0.029 (0.011) & 2.71 \\
n=60, p\_out=0.05, informative & 0.000 (0.000) & 1.00 & 1/99/0 & -0.0000 (0.0001) & -0.01 & 0.000 (0.000) [10] & -- & 0.022 (0.014) & 1.53 \\
n=30, p\_out=0.10, informative & 0.005 (0.007) & 0.64 & 18/66/16 & -0.0002 (0.0007) & -0.26 & 0.014 (0.015) [82] & 0.93 & 0.039 (0.010) & 3.74 \\
n=60, p\_out=0.10, informative & 0.001 (0.002) & 0.53 & 7/89/4 & -0.0000 (0.0002) & -0.08 & 0.018 (0.029) [36] & 0.63 & 0.055 (0.010) & 5.49 \\
n=30, p\_out=0.05, uninformative & 0.003 (0.003) & 0.88 & 4/92/4 & 0.0002 (0.0004) & 0.42 & 0.021 (0.025) [56] & 0.84 & 0.067 (0.014) & 4.73 \\
n=60, p\_out=0.05, uninformative & 0.000 (0.000) & -- & 0/100/0 & 0.0000 (0.0000) & 0.16 & -0.004 (0.004) [4] & -1.00 & 0.014 (0.012) & 1.14 \\
n=30, p\_out=0.10, uninformative & -0.010 (0.008) & -1.27 & 10/71/19 & 0.0002 (0.0008) & 0.22 & 0.015 (0.011) [90] & 1.35 & 0.032 (0.010) & 3.06 \\
n=60, p\_out=0.10, uninformative & -0.002 (0.001) & -1.49 & 1/94/5 & 0.0005 (0.0002) & 3.26 & 0.025 (0.019) [34] & 1.34 & 0.058 (0.009) & 6.52 \\
\bottomrule
\end{tabular}

%% file: tables/Table_B3_Paired_FCoL_vs_PerturbTie.tex
\begin{tabular}{llllllllll}
\toprule
scenario & ARI & ARI z & win/tie/loss & Brier & Brier z & ambiguity AUC & AUC z & structural Spearman & Spearman z \\
\midrule
n=30, p\_out=0.05, informative & -0.001 (0.004) & -0.34 & 5/88/7 & 0.0002 (0.0004) & 0.41 & 0.022 (0.020) [53] & 1.11 & 0.010 (0.009) & 1.19 \\
n=60, p\_out=0.05, informative & -0.000 (0.000) & -0.02 & 1/98/1 & 0.0001 (0.0001) & 1.63 & -0.000 (0.000) [9] & -1.00 & 0.009 (0.012) & 0.74 \\
n=30, p\_out=0.10, informative & 0.003 (0.006) & 0.58 & 15/69/16 & 0.0005 (0.0006) & 0.89 & 0.015 (0.015) [83] & 1.01 & 0.030 (0.008) & 3.65 \\
n=60, p\_out=0.10, informative & 0.001 (0.001) & 0.39 & 4/93/3 & 0.0001 (0.0002) & 0.42 & 0.009 (0.012) [37] & 0.76 & 0.029 (0.010) & 2.97 \\
n=30, p\_out=0.05, uninformative & 0.007 (0.005) & 1.48 & 7/89/4 & 0.0003 (0.0004) & 0.72 & 0.025 (0.018) [55] & 1.42 & 0.041 (0.009) & 4.31 \\
n=60, p\_out=0.05, uninformative & 0.000 (0.000) & -- & 0/100/0 & -0.0000 (0.0000) & -0.23 & -0.004 (0.004) [4] & -1.00 & 0.000 (0.010) & 0.01 \\
n=30, p\_out=0.10, uninformative & -0.012 (0.005) & -2.47 & 9/69/22 & 0.0021 (0.0007) & 3.18 & -0.006 (0.008) [91] & -0.75 & 0.005 (0.007) & 0.63 \\
n=60, p\_out=0.10, uninformative & -0.000 (0.001) & -0.40 & 2/94/4 & 0.0005 (0.0002) & 2.91 & 0.010 (0.011) [35] & 0.92 & 0.032 (0.007) & 4.57 \\
\bottomrule
\end{tabular}

%% file: tables/Table_B4_Asymmetric_Summary.tex
\begin{tabular}{lllllllll}
\toprule
scenario & arm & method & ARI & K\_correct & Brier & structural Spearman & AUC misclassified & share asymmetric \\
\midrule
n=30, p\_out=0.05 & informative & FCo-L & 0.875 (0.012) & 0.69 & 0.0421 (0.0040) & 0.285 (0.019) & 0.804 [52] & 0.80 \\
n=30, p\_out=0.05 & informative & Perturb-core & 0.872 (0.013) & 0.73 & 0.0455 (0.0043) & 0.275 (0.019) & 0.783 [52] & 0.80 \\
n=30, p\_out=0.05 & informative & Perturb-tie & 0.874 (0.013) & 0.72 & 0.0446 (0.0042) & 0.271 (0.021) & 0.793 [53] & 0.80 \\
n=30, p\_out=0.05 & informative & IW-Louvain & 0.897 (0.012) & 0.82 & 0.0435 (0.0049) & -- & -- & 0.80 \\
n=30, p\_out=0.05 & random\_direction & FCo-L & 0.869 (0.013) & 0.71 & 0.0452 (0.0043) & 0.274 (0.022) & 0.825 [53] & 0.79 \\
n=30, p\_out=0.05 & random\_direction & Perturb-core & 0.875 (0.012) & 0.72 & 0.0452 (0.0043) & 0.271 (0.019) & 0.775 [52] & 0.79 \\
n=30, p\_out=0.05 & random\_direction & Perturb-tie & 0.875 (0.013) & 0.73 & 0.0444 (0.0042) & 0.265 (0.020) & 0.787 [52] & 0.79 \\
n=30, p\_out=0.05 & random\_direction & IW-Louvain & 0.870 (0.013) & 0.79 & 0.0547 (0.0053) & -- & -- & 0.79 \\
n=30, p\_out=0.05 & common & Consensus-core & 0.862 (0.015) & 0.71 & 0.0476 (0.0046) & 0.222 (0.021) & 0.747 [53] & -- \\
n=30, p\_out=0.05 & common & Louvain-core & 0.855 (0.014) & 0.68 & 0.0599 (0.0057) & -- & -- & -- \\
n=30, p\_out=0.10 & informative & FCo-L & 0.716 (0.018) & 0.38 & 0.0912 (0.0052) & 0.347 (0.019) & 0.811 [83] & 0.80 \\
n=30, p\_out=0.10 & informative & Perturb-core & 0.703 (0.018) & 0.39 & 0.0976 (0.0055) & 0.345 (0.019) & 0.794 [84] & 0.80 \\
n=30, p\_out=0.10 & informative & Perturb-tie & 0.699 (0.018) & 0.39 & 0.0955 (0.0052) & 0.353 (0.019) & 0.814 [88] & 0.80 \\
n=30, p\_out=0.10 & informative & IW-Louvain & 0.754 (0.018) & 0.84 & 0.1051 (0.0076) & -- & -- & 0.80 \\
n=30, p\_out=0.10 & random\_direction & FCo-L & 0.710 (0.018) & 0.41 & 0.0974 (0.0054) & 0.327 (0.019) & 0.798 [85] & 0.79 \\
n=30, p\_out=0.10 & random\_direction & Perturb-core & 0.700 (0.019) & 0.39 & 0.0985 (0.0055) & 0.328 (0.019) & 0.799 [84] & 0.79 \\
n=30, p\_out=0.10 & random\_direction & Perturb-tie & 0.698 (0.018) & 0.41 & 0.0968 (0.0054) & 0.335 (0.021) & 0.795 [88] & 0.79 \\
n=30, p\_out=0.10 & random\_direction & IW-Louvain & 0.717 (0.018) & 0.76 & 0.1192 (0.0076) & -- & -- & 0.79 \\
n=30, p\_out=0.10 & common & Consensus-core & 0.696 (0.018) & 0.35 & 0.1014 (0.0057) & 0.300 (0.020) & 0.767 [86] & -- \\
n=30, p\_out=0.10 & common & Louvain-core & 0.673 (0.019) & 0.38 & 0.1328 (0.0075) & -- & -- & -- \\
n=60, p\_out=0.05 & informative & FCo-L & 0.998 (0.001) & 0.99 & 0.0011 (0.0004) & 0.061 (0.019) & 0.585 [4] & 0.80 \\
n=60, p\_out=0.05 & informative & Perturb-core & 0.995 (0.002) & 0.98 & 0.0015 (0.0004) & 0.088 (0.016) & 0.763 [7] & 0.80 \\
n=60, p\_out=0.05 & informative & Perturb-tie & 0.996 (0.001) & 0.97 & 0.0014 (0.0004) & 0.077 (0.016) & 0.723 [6] & 0.80 \\
n=60, p\_out=0.05 & informative & IW-Louvain & 0.997 (0.001) & 1.00 & 0.0013 (0.0005) & -- & -- & 0.80 \\
n=60, p\_out=0.05 & random\_direction & FCo-L & 0.995 (0.001) & 1.00 & 0.0016 (0.0005) & 0.083 (0.016) & 0.815 [9] & 0.78 \\
n=60, p\_out=0.05 & random\_direction & Perturb-core & 0.996 (0.001) & 1.00 & 0.0015 (0.0004) & 0.068 (0.017) & 0.644 [7] & 0.78 \\
n=60, p\_out=0.05 & random\_direction & Perturb-tie & 0.996 (0.001) & 1.00 & 0.0014 (0.0004) & 0.074 (0.017) & 0.792 [8] & 0.78 \\
n=60, p\_out=0.05 & random\_direction & IW-Louvain & 0.994 (0.002) & 1.00 & 0.0026 (0.0007) & -- & -- & 0.78 \\
n=60, p\_out=0.05 & common & Consensus-core & 0.996 (0.001) & 1.00 & 0.0015 (0.0004) & 0.084 (0.017) & 0.792 [8] & -- \\
n=60, p\_out=0.05 & common & Louvain-core & 0.991 (0.002) & 1.00 & 0.0039 (0.0010) & -- & -- & -- \\
n=60, p\_out=0.10 & informative & FCo-L & 0.973 (0.004) & 1.00 & 0.0087 (0.0012) & 0.250 (0.014) & 0.944 [42] & 0.80 \\
n=60, p\_out=0.10 & informative & Perturb-core & 0.969 (0.004) & 1.00 & 0.0106 (0.0014) & 0.267 (0.014) & 0.960 [44] & 0.80 \\
n=60, p\_out=0.10 & informative & Perturb-tie & 0.971 (0.004) & 1.00 & 0.0104 (0.0013) & 0.257 (0.014) & 0.953 [44] & 0.80 \\
n=60, p\_out=0.10 & informative & IW-Louvain & 0.979 (0.003) & 1.00 & 0.0090 (0.0013) & -- & -- & 0.80 \\
n=60, p\_out=0.10 & random\_direction & FCo-L & 0.970 (0.004) & 1.00 & 0.0103 (0.0014) & 0.264 (0.015) & 0.950 [43] & 0.79 \\
n=60, p\_out=0.10 & random\_direction & Perturb-core & 0.968 (0.005) & 1.00 & 0.0105 (0.0014) & 0.267 (0.014) & 0.948 [45] & 0.79 \\
n=60, p\_out=0.10 & random\_direction & Perturb-tie & 0.970 (0.004) & 1.00 & 0.0105 (0.0014) & 0.267 (0.015) & 0.932 [44] & 0.79 \\
n=60, p\_out=0.10 & random\_direction & IW-Louvain & 0.975 (0.004) & 1.00 & 0.0108 (0.0017) & -- & -- & 0.79 \\
n=60, p\_out=0.10 & common & Consensus-core & 0.968 (0.004) & 1.00 & 0.0109 (0.0014) & 0.253 (0.014) & 0.958 [46] & -- \\
n=60, p\_out=0.10 & common & Louvain-core & 0.949 (0.006) & 0.99 & 0.0222 (0.0027) & -- & -- & -- \\
\bottomrule
\end{tabular}

%% file: tables/Table_B5_Asymmetric_Paired.tex
\begin{tabular}{lllllllll}
\toprule
scenario & arm & comparison & ARI & ARI z & Brier & Brier z & structural Spearman & Spearman z \\
\midrule
n=30, p\_out=0.05 & informative & FCo-L minus Consensus-core & 0.012 (0.006) & 2.25 & -0.0056 (0.0013) & -4.24 & 0.062 (0.014) & 4.55 \\
n=60, p\_out=0.05 & informative & FCo-L minus Consensus-core & 0.002 (0.001) & 1.97 & -0.0004 (0.0002) & -2.16 & -0.010 (0.011) & -0.90 \\
n=30, p\_out=0.10 & informative & FCo-L minus Consensus-core & 0.020 (0.008) & 2.58 & -0.0102 (0.0016) & -6.29 & 0.047 (0.010) & 4.69 \\
n=60, p\_out=0.10 & informative & FCo-L minus Consensus-core & 0.005 (0.002) & 2.70 & -0.0023 (0.0004) & -6.29 & -0.008 (0.009) & -0.90 \\
n=30, p\_out=0.05 & random\_direction & FCo-L minus Consensus-core & 0.006 (0.006) & 1.17 & -0.0024 (0.0012) & -2.08 & 0.058 (0.015) & 3.79 \\
n=60, p\_out=0.05 & random\_direction & FCo-L minus Consensus-core & -0.001 (0.001) & -1.00 & 0.0001 (0.0001) & 0.71 & 0.013 (0.012) & 1.09 \\
n=30, p\_out=0.10 & random\_direction & FCo-L minus Consensus-core & 0.014 (0.007) & 2.08 & -0.0040 (0.0012) & -3.42 & 0.027 (0.010) & 2.64 \\
n=60, p\_out=0.10 & random\_direction & FCo-L minus Consensus-core & 0.002 (0.002) & 0.81 & -0.0006 (0.0003) & -2.34 & 0.009 (0.010) & 0.85 \\
n=30, p\_out=0.05 & informative & FCo-L minus Perturb-core & 0.003 (0.004) & 0.67 & -0.0034 (0.0009) & -3.98 & 0.003 (0.007) & 0.42 \\
n=60, p\_out=0.05 & informative & FCo-L minus Perturb-core & 0.002 (0.001) & 1.83 & -0.0004 (0.0002) & -2.55 & -0.029 (0.015) & -1.89 \\
n=30, p\_out=0.10 & informative & FCo-L minus Perturb-core & 0.013 (0.008) & 1.65 & -0.0063 (0.0013) & -5.06 & 0.002 (0.012) & 0.19 \\
n=60, p\_out=0.10 & informative & FCo-L minus Perturb-core & 0.004 (0.002) & 2.62 & -0.0019 (0.0003) & -6.24 & -0.023 (0.009) & -2.45 \\
n=30, p\_out=0.05 & random\_direction & FCo-L minus Perturb-core & -0.007 (0.004) & -1.78 & 0.0000 (0.0008) & 0.02 & 0.000 (0.015) & 0.01 \\
n=60, p\_out=0.05 & random\_direction & FCo-L minus Perturb-core & -0.001 (0.001) & -1.42 & 0.0001 (0.0001) & 1.28 & 0.017 (0.013) & 1.36 \\
n=30, p\_out=0.10 & random\_direction & FCo-L minus Perturb-core & 0.010 (0.006) & 1.66 & -0.0011 (0.0011) & -1.01 & -0.002 (0.010) & -0.18 \\
n=60, p\_out=0.10 & random\_direction & FCo-L minus Perturb-core & 0.002 (0.002) & 0.99 & -0.0002 (0.0002) & -1.10 & -0.006 (0.009) & -0.61 \\
n=30, p\_out=0.05 & informative & FCo-L minus Perturb-tie & 0.001 (0.004) & 0.18 & -0.0025 (0.0008) & -3.09 & -0.002 (0.009) & -0.18 \\
n=60, p\_out=0.05 & informative & FCo-L minus Perturb-tie & 0.002 (0.001) & 1.92 & -0.0003 (0.0001) & -2.07 & -0.014 (0.006) & -2.25 \\
n=30, p\_out=0.10 & informative & FCo-L minus Perturb-tie & 0.017 (0.006) & 2.97 & -0.0043 (0.0010) & -4.48 & -0.006 (0.010) & -0.62 \\
n=60, p\_out=0.10 & informative & FCo-L minus Perturb-tie & 0.002 (0.001) & 2.03 & -0.0018 (0.0003) & -6.48 & -0.012 (0.009) & -1.32 \\
n=30, p\_out=0.05 & random\_direction & FCo-L minus Perturb-tie & -0.006 (0.004) & -1.74 & 0.0008 (0.0008) & 1.00 & 0.005 (0.009) & 0.52 \\
n=60, p\_out=0.05 & random\_direction & FCo-L minus Perturb-tie & -0.001 (0.001) & -1.00 & 0.0001 (0.0001) & 1.37 & 0.009 (0.013) & 0.74 \\
n=30, p\_out=0.10 & random\_direction & FCo-L minus Perturb-tie & 0.012 (0.006) & 2.12 & 0.0006 (0.0009) & 0.73 & -0.008 (0.009) & -0.86 \\
n=60, p\_out=0.10 & random\_direction & FCo-L minus Perturb-tie & -0.000 (0.002) & -0.06 & -0.0002 (0.0002) & -0.96 & -0.012 (0.010) & -1.25 \\
n=30, p\_out=0.05 & informative & FCo-L minus IW-Louvain & -0.022 (0.008) & -2.85 & -0.0015 (0.0026) & -0.55 & -- (NA) & -- \\
n=60, p\_out=0.05 & informative & FCo-L minus IW-Louvain & 0.001 (0.001) & 1.01 & -0.0002 (0.0003) & -0.66 & -- (NA) & -- \\
n=30, p\_out=0.10 & informative & FCo-L minus IW-Louvain & -0.037 (0.014) & -2.67 & -0.0138 (0.0051) & -2.70 & -- (NA) & -- \\
n=60, p\_out=0.10 & informative & FCo-L minus IW-Louvain & -0.007 (0.003) & -2.44 & -0.0004 (0.0008) & -0.45 & -- (NA) & -- \\
n=30, p\_out=0.05 & random\_direction & FCo-L minus IW-Louvain & -0.002 (0.009) & -0.18 & -0.0095 (0.0034) & -2.75 & -- (NA) & -- \\
n=60, p\_out=0.05 & random\_direction & FCo-L minus IW-Louvain & 0.002 (0.001) & 1.14 & -0.0011 (0.0005) & -2.06 & -- (NA) & -- \\
n=30, p\_out=0.10 & random\_direction & FCo-L minus IW-Louvain & -0.007 (0.012) & -0.53 & -0.0218 (0.0047) & -4.66 & -- (NA) & -- \\
n=60, p\_out=0.10 & random\_direction & FCo-L minus IW-Louvain & -0.005 (0.003) & -1.92 & -0.0005 (0.0010) & -0.51 & -- (NA) & -- \\
n=30, p\_out=0.05 & between arms & FCo-L informative minus FCo-L random direction & 0.006 (0.004) & 1.34 & -0.0031 (0.0008) & -3.70 & 0.002 (0.014) & 0.15 \\
n=60, p\_out=0.05 & between arms & FCo-L informative minus FCo-L random direction & 0.002 (0.001) & 2.23 & -0.0005 (0.0002) & -2.78 & -0.021 (0.012) & -1.68 \\
n=30, p\_out=0.10 & between arms & FCo-L informative minus FCo-L random direction & 0.006 (0.007) & 0.92 & -0.0062 (0.0013) & -4.92 & 0.021 (0.012) & 1.69 \\
n=60, p\_out=0.10 & between arms & FCo-L informative minus FCo-L random direction & 0.003 (0.002) & 1.71 & -0.0017 (0.0003) & -5.42 & -0.016 (0.009) & -1.78 \\
n=30, p\_out=0.05 & between arms & Perturb-tie informative minus Perturb-tie random direction & -0.001 (0.003) & -0.34 & 0.0001 (0.0003) & 0.49 & 0.010 (0.011) & 0.92 \\
n=60, p\_out=0.05 & between arms & Perturb-tie informative minus Perturb-tie random direction & 0.000 (0.001) & 0.40 & -0.0001 (0.0001) & -0.78 & 0.003 (0.010) & 0.26 \\
n=30, p\_out=0.10 & between arms & Perturb-tie informative minus Perturb-tie random direction & 0.001 (0.005) & 0.18 & -0.0013 (0.0006) & -2.03 & 0.018 (0.009) & 1.96 \\
n=60, p\_out=0.10 & between arms & Perturb-tie informative minus Perturb-tie random direction & 0.001 (0.001) & 0.79 & -0.0001 (0.0002) & -0.45 & -0.014 (0.008) & -1.79 \\
\bottomrule
\end{tabular}